\documentclass[10pt,onecolumn,twoside]{IEEEtran} 

\usepackage{multirow}
\usepackage{amsfonts} 
\usepackage{epsfig} 
\usepackage{amsmath} 
\usepackage{amssymb}
\usepackage[nolist]{acronym}
\usepackage[english]{babel}
\usepackage{color} 
\usepackage{stfloats}  
\usepackage{algorithm}
\usepackage{algorithmic}
\usepackage{subfigure}
\usepackage{pgfplots}
\usepackage{pgfplotstable}
\usepackage{caption}
\usepackage{booktabs}
\usepackage{tabularx}
\usepackage{url}
\usetikzlibrary{shapes,positioning}
\usetikzlibrary{decorations,decorations.pathmorphing}
\usepackage{wrapfig}

\newcommand{\mat}{\pmb}

\newcommand{\excess}{ downloading rate overshooting}

\newfloat{ilp}{ht}{aux}
\floatname{ilp}{Integer Linear Program}

\begin{document}

\title{Optimized Adaptive Streaming Representations based on System Dynamics}
 
\author{ Laura  Toni, Ramon Aparicio-Pardo, Karine Pires,  Gwendal Simon, Alberto Blanc, and Pascal Frossard
 \thanks{L. Toni,  and P. Frossard are with \'Ecole Polytechnique F\'ed\'erale de Lausanne (EPFL), Signal Processing Laboratory - LTS4, CH-1015 Lausanne, Switzerland. Email: \{laura.toni, pascal.frossard\}@epfl.ch.} \\
 \thanks{R. Aparicio-Pardo, K. Pires, G. Simon, and A. Blanc, D\'ept. R\'eseaux, S\'ecurit\'e et Multim\'edia, Technopole Brest-Iroise - CS 83818 - 29238 Brest Cedex 3, France; email: \{ramon.aparicio, karine.pires, gwendal.simon, alberto.blanc\}@telecom-bretagne.eu. }
 \thanks{This work was partially funded by the Swiss National Science Foundation (SNSF) under the CHIST-ERA project CONCERT (A Context-Adaptive Content Ecosystem Under Uncertainty)}
}

\begin{acronym}
\acro{DASH}{dynamic adaptive streaming over HTTP}  
\acro{CDN}{content delivery network}  
\acro{CDF}{cumulative density function}
\acro{PDF}{probability density function}  
\acro{RTP}{real-time protocol}
\acro{QoE}{Quality of Experience}
\acro{VQM}{Video Quality Metric}
\acro{MILP}{mixed integer linear program}
\acro{ILP}{integer linear program}
\acro{HDTV}{high definition television}
\acro{UGC}{User-Generated Content}
\acro{MPD}{Media Presentation Description}
\end{acronym}

 \maketitle  
\begin{abstract}
Adaptive streaming addresses the increasing and heterogeneous demand of 
multimedia content over the Internet by offering several encoded versions for 
each video sequence. Each version (or representation) is characterized by a 
resolution and a bit rate, and it is aimed at a specific set of users,  like TV 
or mobile phone clients.  While most existing works on adaptive streaming deal 
with effective playout-buffer control strategies on the client side, in this 
paper we take  a providers' perspective and propose solutions to improve user 
satisfaction by optimizing the set of available representations.  We formulate 
an integer linear program   that maximizes users' average satisfaction, taking 
into account network dynamics, type of video content, and user population 
characteristics.  The solution of the optimization is a set of encoding 
parameters corresponding to the representations set that maximizes user 
satisfaction.
We evaluate this solution by simulating multiple adaptive streaming sessions 
characterized by realistic network   statistics,  showing that  the 
proposed solution  outperforms  commonly used  vendor recommendations, in terms 
of user satisfaction but also in terms of fairness and outage probability.  The 
simulation results show  that video content information as well as network 
constraints and users'   statistics play a crucial role in selecting 
proper encoding parameters to provide fairness among users and to reduce 
network resource usage.   We finally  propose a few theoretical  guidelines that can 
be used, in realistic settings, to choose  the encoding parameters based on the 
user  characteristics, the network capacity and the type of video content. 
\end{abstract}

\begin{keywords}
Dynamic adaptive streaming over HTTP, content distribution, video streaming, integer linear program.
\end{keywords}

\maketitle

\section{Introduction}
\begin{figure}[t]
\begin{center}
\includegraphics[width=0.5\linewidth,draft=false]{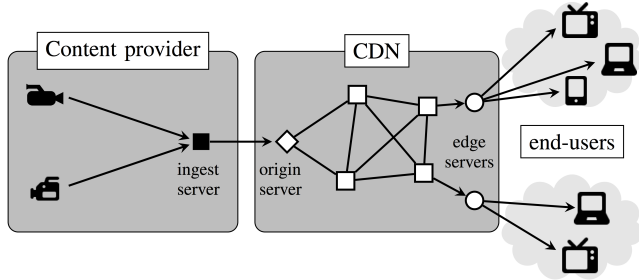}
\caption{Video content  delivery chain.}\label{cdn}
\end{center}
\end{figure}

Due to the ever increasing popularity of modern mobile devices, users can 
request and play  multimedia content anywhere and at any time. This results in an 
increase of the variety of each of the following: requested contents, devices 
used to display them and access network capacity~\cite{videotraffic}.  
Adaptive streaming solutions aim at addressing this growing heterogeneity  by 
offering several versions of the video sequences. Each version is  encoded at a 
different bitrate and resolution so that each user can select the  most suitable 
version depending on the video client capabilities and network bandwidth. 
Fig.~\ref{cdn} illustrates an instance of an adaptive streaming system. The 
ingest server receives video data from cameras and prepares several different 
video \emph{representations},  each one characterized by a different resolution 
and bitrate. The ingest server sends the streams corresponding to each 
representation to the origin server of a~\ac{CDN}, which delivers the video 
representations to the edge-servers, which, in turn, directly serve the requests 
of the clients. 

Several models have been  recently proposed to standardize the adaptive 
streaming communication framework, like 
\ac{DASH}~\cite{MPEG-DASH:12,stockhammer2011,sodagar2011mpeg} and 
WebRTC~\cite{WebRTC}. The multiple  implementations of such systems differ in 
two ways: $(i)$ the client adaptation strategy, and $(ii)$ the selection of the 
different video representations. So far, the first problem has been at the 
center of the attention of the research community, while the second one has 
rarely been considered. The only existing guidelines  for selecting the 
parameters of the video representations are \emph{recommendations} from system 
manufacturers, including Apple~\cite{apple-setting} and  
Microsoft~\cite{SmoothStr}. Some content providers have also defined their own  
representations sets, for example Netflix~\cite{netFlix-setting}.  However, to 
the best of our knowledge, neither the recommendations from system manufacturers 
nor the choices made by content providers have been supported by any scientific 
study.

This paper is a first step towards filling this gap.  We focus  on optimizing  
the \emph{set of representations} that should be generated by the ingest server 
and show that the existing recommended sets have critical  weaknesses. 
Optimizing the encoding parameters for representations sets is an open problem,  
dealing with multiple  constraints, including the cost of delivering  video 
streams using a \ac{CDN}, the characteristics of end-users, and the type  of 
video to be delivered. For example, smaller sets (i.e., with few representations 
for  each video) might satisfy only a fraction of the users, while larger ones 
could  satisfy more users, but at a larger cost  in terms of increased storage 
costs  for on-demand video, or larger encoding delays in the case of live 
streaming. It is therefore important to study how the representations set should 
be designed, in order to strike the appropriate balance between user 
satisfaction and the cost of the system. This is the goal of our work. 

In particular, we consider a scenario in which video channels (e.g., sports, 
documentary, cartoon) are encoded at different encoding rates and various 
spatial resolutions, leading to  several  representations available at the 
server. Representations are then delivered to clients through a  \ac{CDN} 
characterized by  an overall capacity constraint.  Clients requests (video 
content and display resolution) are supposed to be known. Depending on the 
bandwidth  available to each user, specific representations will be provided to 
fulfill clients' video requests. The quality experienced by users, modeled in 
our problem as a satisfaction level function, depends on   both the compression 
artifacts (driven by the video source rate) and spatial scaling  artifacts 
(depending on the potential adaptation of the video resolution to the display 
resolution). 

We formulate an optimization problem to select   the best encoding parameters of 
the representations set and study the resulting performance in different adaptive 
streaming scenarios.  We further show the need of making the selection of the 
representations set based on the video content, network, and clients' 
characteristics.   The proposed optimization is not necessarily meant to be  used to make real-time adjustments in DASH systems. It is rather a theoretical framework to derive benchmarks  or optimal encoding solutions for non-live   systems, along with guidelines for practical design of representations sets in video streaming applications.  
 The provided optimization   highlights  the sub-optimality of the current recommendation sets and  provide theoretical guidelines that can help a system designer to understand  which are the crucial system parameters that should be taken into account when  optimizing the representations set.  Our main contributions are as follows:

\begin{description}
\item[i.] We formulate a novel \ac{ILP} to find the best representations set, 
defined as the one that maximizes the  expected user satisfaction under network 
and system constraints. The  satisfaction of each client is a function of the 
encoding rate, the resolution, the characteristics of the requested video, and 
of the bandwidth that can be used to deliver the video. By using a generic 
solver, it is  possible to solve the \ac{ILP} on representative cases, gaining 
insights about the optimal representations sets.

\vspace{2mm}

\item[ii.] We use the \ac{ILP} to study the recommendations from system 
manufacturers and content providers.   We compute the solution of the \ac{ILP} 
for different user populations and study how  it performs in realistic streaming 
applications w.r.t. the existing recommendations. Simulation results show 
that recommended sets lead to good performance in terms of quality experienced 
by those users that can be served by the system, but   they also lead to a large 
probability for users not to be properly served. Overall, recommended sets require  too many 
representations, do not easily adapt to system dynamics, and lead to unfair 
sharing of the network among users.   

\vspace{2mm}

\item[iii.]  In order to provide insights on how a system provider should select 
the  encoding rates sets, we study the optimal representations sets in 
different scenarios.  We consider several realistic cases, by varying key parameters like the 
number of users requesting each resolution, network connection (capacity of each 
client connection and overall CDN capacity), and type of video (sport, 
documentary, movie, cartoon). 
By analyzing the solution of the \ac{ILP} in each scenario, we notice 
recurrent patterns and derive   few generic \emph{guidelines}, which 
can be useful for content providers in the selection of the best encoding 
parameters. 
\end{description}

The remainder of this paper is organized as follows. Related works on adaptive 
streaming are described in Section~\ref{sec:rel_work}. The formulation of the 
optimization problem as an \ac{ILP} is provided in Section~\ref{sec:probl_form}. 
In Section~\ref{sec:evaluation-settings}, we detail the simulation settings. In 
Section~\ref{sec:numerical-analysis}, results are provided to study the system 
performance of optimal representations sets  w.r.t the recommended one. In 
Section~\ref{sec:guidelines}, we   analyze the behavior of the 
optimal set across different configuration to derive the guidelines. Finally, 
conclusions and future works are discussed in Section~\ref{sec:concl}.

\section{Related Works}
\label{sec:rel_work}
During the last decade, adaptive streaming has been an active research area, 
with most efforts aimed at developing server-controlled streaming solutions.  
Recently, a client-driven approach, based  on HTTP-adaptive 
streaming~\cite{stockhammer2011,sodagar2011mpeg}, has gained popularity and 
attention.  In this new paradigm, the clients  decide which segments to get and 
when to request them, and the server mainly responds to the clients' requests. 
Different implementation of this new architecture have been proposed in various 
commercial DASH players~\cite{Akhshabi2012271}. 

Most of the research effort in adaptive streaming  has been devoted to   
improve the client controller, i.e., to optimize the  representations selection 
for each user~\cite{DeCicco_Masc:J14,Cong_Thang:JSAC14,miller2012}. The 
controller behavior is generally driven by an estimate of the network 
dynamics~\cite{Begen_DASH:2013} and the state of the client 
buffer~\cite{Huang_Johari:ARXIV14}. The general objective is to maximize the 
\ac{QoE} for the users while avoiding unnecessary quality fluctuations. For 
example, the selection of the representation can be optimized in such a way 
that large variations of rates in successive segments are avoided, since large 
rate variations may lead to an unpleasant viewing 
experience~\cite{Li_Begen:ArXiv14,Veciana:2013,Mok:2012}. Other solutions for 
the controller have also been investigated in order to minimize the re-buffering 
phases~\cite{miller2012,Li_Begen:ArXiv14}. On a more general perspective, it has 
been shown that the current HTTP-adaptive streaming systems have limitations 
when a  large number of clients share the same network~\cite{Akhshabi:2012}. 
Hence, some recent research works modify the client controllers in order to 
simultaneously reach fairness and efficiency when many clients share the same 
bottleneck link~\cite{Jiang:J14,Zhu_Li:MMSP13}.  Rather than focusing on the  client controller design, the work in  \cite{Essaili:C13} investigates a \ac{DASH} streaming  system over a mobile network where a proxy rewrites client HTTP requests in such  a way that the overall \ac{QoE} experienced by multiple clients is optimized.  
The work in~\cite{Essaili:C13}  addresses the main limitations of   
multiple-clients \ac{DASH} systems; however
it does not address the problem of optimizing the representations on the server and rather seems complementary to our work.  

Despite the many, recently published, papers about \ac{DASH}, the problem of 
selecting proper representations to be stored on the server has been mostly 
overlooked.  The set of available representations is usually supposed to be 
known (and fixed). These representations are often selected based on vendor  or 
content provider recommendations, as in the case of Apple~\cite{apple-setting}, 
 Microsoft~\cite{SmoothStr}, and Netflix~\cite{netFlix-setting}.  To the best 
of our knowledge neither the recommendations from system manufacturers nor the 
choices made by content providers have been supported by any scientific study in 
the literature. Rather they seem to be based on admittedly fairly good 
heuristics.
 
The importance of the optimized representations sets in adaptive streaming has 
recently been highlighted in~\cite{Cong_Thang:JSAC14}, where  authors show  that 
the representations sets may affect the behaviors of some 
adaptation methods. For example, a gain can be achieved when the 
representations set available at the server is selected based on the video 
content information rather than simply the rate information. However, the 
authors do not propose an optimization of the set nor guidelines on  the  
selection of the representations set.  Encoding rate optimization has been 
investigated very recently in~\cite{Zahn:13}, for on-demand videos in a 
storage-limited scenario. Rates are  optimized in such a way that the best 
possible \ac{QoE} is provided to a pool of users and a total storage capacity 
constraint is met. All the scenarios  presented in~\cite{Zahn:13} consider a 
homogeneous user population and this is a key  assumption exploited in the 
solution of the optimization problem.  In~\cite{Toni:C14}, the optimization of 
the set of representations in the case where heterogeneous users are 
characterized by a static link capacity and a single acceptable resolution has 
been studied.
In this paper, instead, we explicitly model  different types of users, in terms 
of access link capacity and devices used. We also take into account the dynamic 
aspect of the channel as well as different types of video as this has a 
non-negligible impact on the perceived \ac{QoE}. 


\section{Problem Formulation}
\label{sec:probl_form}
We now present the problem formulation.  The goal is to select the  best 
representations set,  taking into account video content, available network 
capacity and users' characteristics. We consider the user population (in terms 
of requested video content and resolution) and the CDN total capacity 
as known values.  We model the time-varying available capacity between the CDN 
and each client using one \ac{CDF} for each client. Statistics are extrapolated from  the publicly 
available dataset  presented in~\cite{Basso:C14},  where  network 
measurements have been collected by a DASH module from more than a thousand 
Internet clients.     


In the following, we first introduce the notation used in the 
problem formulation. Then we present the \ac{ILP}  used to compute the optimal 
representations sets.

\subsection{Definitions}
Let $\mathcal{V}$ be the set of videos. Each video $v\in \mathcal{V}$ can be 
encoded using different representations, each one characterized by the encoding 
rate $r\in \mathcal{R}$ and the spatial resolution $s\in \mathcal{S}$, being 
$\mathcal{R}$ and  $\mathcal{S}$ respectively the sets of bit rates and spatial 
resolutions that are admissible for the representations. The triple ($v$, $r$, 
$s$) corresponds to the representation of a video $v\in \mathcal{V}$ encoded  at    rate $r \in \mathcal{R}$ and resolution $s \in \mathcal{S}$. Note that   $r$ is a pure integer number that represents  the rate index in $\mathcal{R}$. The nominal value (in  $kbps$) of the encoding rate $r$ is denoted by $b_r$. Each 
resolution $s$ admits encoding rates within the range $[b_{vs}^{\text{min}}$,  
$b_{vs}^{\text{max}}]$ for video $v$.

Let $\mathcal{U}$ be the set of users that the CDN network should serve, where 
each user $u \in \mathcal{U}$ requests a video $v_u \in \mathcal{V}$ and plays 
the video representation at a given spatial resolution $s_u \in \mathcal{S}$ 
corresponding to the user display resolution (i.e., the spatial resolution at 
which the video will be displayed on the user's device). We follow the 
assumption that a user $u$ can play segments encoded at resolutions different 
from its display size by performing spatial down-sampling/up-sampling before 
rendering. We denote by $T_{ur}$ the  percentage of time that user $u$ has a link capacity larger than 
 $b_r$ for a certain encoded rate $r$.  These parameters are computed from the  cumulative distribution 
function of the measured throughput of the user $u$, using the dataset 
described in~\cite{Basso:C14}

\begin{table}[t]
\centering
\scriptsize
\begin{tabularx}{\textwidth }{ >{} lX }
\toprule
Name & Description \\
\midrule
 $f_{uvrs} \in \mathcal {R}^+$ & Satisfaction level for the representation of video $v$, watched at display $s_u$ and encoded at rate $r$ and resolution $s$ \\
 $T_{ur} \in [0,1]$ & Percentage of time during  which the throughput of user $u$ is larger than the value $b_r$ of the encoding rate $r$\\
 $T_{\text{min}} \in [0,1]$ & Minimum percentage of time during which a user is served\\
 $b_r \in \mathcal {R}^+$ &Value in $kbps$ of the encoding  rate $r$ \\
 $b_{vs}^{\text{min}} \in \mathcal {R}^+$ &Value in $kbps$ of the minimum 
encoding rate that the video $v$ at resolution $s$ can admit. \\
 $b_{vs}^{\text{max}} \in \mathcal {R}^+$ &Value in $kbps$ of the maximum 
encoding rate  that the video $v$ at resolution $s$ can admit. \\
 $v_u \in \mathcal{V}$ &  Video channel requested by user $u$ \\
 $s_u \in \mathcal{S}$ &  Display size  (spatial resolution) for user $u$ \\
$C \in \mathcal {R}^+$ & Average CDN budget defined as average capacity per user in $kbps$\\
$K \in \mathcal {R}^+$ & Total number of representations used, i.e., triples 
($v$, $r$, $s$) available at the server\\
$P \in [0,1]$ & Fraction of users that must be served \\
\bottomrule	
\end{tabularx}
\caption{Notation adopted in the \ac{ILP} formulation.}\label{tab:notation}
\end{table}

A user $u$ with a  display resolution $s_u$ watching video $v$ encoded at 
resolution $s$ experiences a satisfaction level $f_{uvs}(r)\in [0,1]$, which 
is an increasing function of the encoding rate $r$. Generally, for a given pair ($v$, 
$r$), the satisfaction level is higher if the video resolution $s$ is the same 
as the display resolution $s_u$ than if $s\neq s_u$. This is due to artifacts 
introduced by the up-sampling and down-sampling of the spatial resolution 
during the decoding process on the user side. For the sake of clarity, throughout the paper we denote 
the satisfaction level by $f_{uvrs}$ rather than $f_{uvs}(r)$.
 
We  define  the optimal  encoding parameters  set as the one  that  maximizes 
the expected user satisfaction, subject to several constraints imposed by both 
the delivery system and the service provider. 
The constraints that we formulate 
for this problem  derive directly  from  real challenges  identified by service 
providers. We highlight three  such constraints:
\begin{description}
\item[i.] \textbf{The overall CDN capacity} available to deliver all the video 
streams. In general, video service providers reserve an overall budget (in \$) 
for video delivery and use it to buy a delivery service from a CDN provider.  
In today's CDN, the price depends on the sum of all the rates of all the video 
streams originating at the content provider~\cite{NygrenSS10}. Thus, the  video 
service provider is interested in maintaining the total delivery bandwidth below 
a given value, here represented by $C \cdot |\mathcal{U}|$, where $C$ denotes 
the average CDN budget in terms of hired capacity \emph{per user} in $kbps$ and $ 
|\mathcal{U}|$ denotes the number of users of the CDN. 
\vspace{2mm}
\item[ii.]
\textbf{The total number of representations}, denoted by $K$, is  the total 
number of triplets ($v$, $r$, $s$)   provided to the ingest 
servers. A higher number of representations means more complexity and higher 
system costs for the video service provider. Higher complexity comes from more 
data to handle, log, store and deliver while system cost directly derives from 
the number of machines that have to be provisioned to encode raw video and from 
storage costs. To have an idea of possible storage and maintenance  costs, a 
website like justin.tv has to maintain about $4,\!000$ video  channels 
simultaneously~\cite{justintv}.
\vspace{2mm}
\item[iii.]
\textbf{The minimal fraction of time during which some users should be served}.  
Ideally, the service  provider would like to serve all the users.  But in 
certain cases, especially when the number of representations $K$ is small, users 
might not be served if the channel capacity is too small for the available 
representations.  In this case, the representations set which optimizes the 
average satisfaction might not lead to  fairness among users.   To address this 
problem, we impose that at least a fraction $P$ of users must be served for at 
least a fraction of  time $T_{\text{min}}$. \footnote{As there exist different 
definitions of fairness, this  constraint can be modified accordingly.} 
\end{description}
Table~\ref{tab:notation} summarizes the notation used in this paper.

\subsection{\ac{ILP} Model}
\label{sec:ilp-model}
We now describe the \ac{ILP} formulation for computing the optimal set of 
representations. The decision variables in our framework model are the 
following:

\begin{itemize}
\item \ $\tau_{uvrs}  \in [0,1]$: percentage of time during which user $u$ 
is served by a representation of video $v$ at resolution $s$ and rate $r$ 

\item  \ $\displaystyle\begin{aligned}[t]
\alpha_{uvrs}  = \left\{
	\begin{array}{rl}
		1, & \text{if user $u$ is served by a representation of video 
			   $v$ encoded at resolution $s$}
		 \\ &  \text{and at rate $r$} \\
		0, & \text{otherwise}.
	\end{array} \right.
\end{aligned}$

\item \ $\displaystyle\begin{aligned}[t]
\beta_{vrs} = \left\{
	\begin{array}{rl}
		1, & \text{if any user in the system is served by a 
			   representation of video $v$}   \\  
		    & \text{encoded at resolution $s$ and at rate $r$} \\
		0, & \text{otherwise}.
	\end{array} \right.
\end{aligned}$

\item \ $\displaystyle\begin{aligned}[t]
\gamma_{u} = \left\{
	\begin{array}{rl}
		1, & \text{if a user $u$ is served by any video representation} \\
		0, & \text{otherwise}.
	\end{array} \right.
\end{aligned}$
 \end{itemize}
With these variables, the optimization problem can be formulated as shown in 
\eqref{formulation_one_resol_resolbased}.
\begin{ilp}[t]
\begin{subequations}\label{formulation_one_resol_resolbased}
\begin{align}     
&\hspace{-0.4cm} \max_{\{\mat{\tau},\mat{\alpha},\mat{\beta},\mat{\gamma}\}} \sum_{u \in \mathcal{U}}  \sum_{v \in \mathcal{V}} \sum_{r \in \mathcal{R}} \sum_{s \in \mathcal{S}} f_{uvrs}  \cdot\tau_{uvrs}        \label{eq:objective}     \\
&\hspace{-0.3cm}\text{s.t.  } 
 \tau_{uvrs} \leq \alpha_{uvrs}, 								&\hspace{-3cm} u \in \mathcal{U},   v \in \mathcal{V},  r \in \mathcal{R}, s \in \mathcal{S} \label{eq:alpha_on} \\
&\hspace{0.1cm} \alpha_{uvrs} \leq \beta_{vrs}, 						& u \in \mathcal{U},   v \in \mathcal{V},  r \in \mathcal{R}, s \in \mathcal{S} \label{eq:beta_off} \\
&\hspace{0.1cm} \beta_{vrs} \leq \sum_{u \in \mathcal{U}} \alpha_{uvrs}, 	 	& v \in \mathcal{V},  r \in \mathcal{R}, s \in \mathcal{S} \label{eq:beta_on}   \\
&\hspace{0.1cm} \sum_{v \in \mathcal{V}} \sum_{s \in \mathcal{S}}  \sum_{\substack{r' \in \mathcal{R} \\ r' \geq r}} \tau_{uvr's} \leq T_{ur}, 	 	& u \in \mathcal{U},  r \in \mathcal{R}  \label{eq:cdf}   \\
&\hspace{0.1cm} \sum_{r \in \mathcal{R}} \sum_{s \in \mathcal{S}}\tau_{uvrs}  \leq  \left\{
	\begin{array}{rl}
		1,&  \text{ if $v = v_u$}\\
			& \text{ \& $s \in \{s_u-1,s_u,s_u+1\} $ }\\
		0,& \text{ otherwise}\\
	\end{array} \right.
&\hspace{-1.2cm} u \in \mathcal{U},   v \in \mathcal{V}   \label{eq:demand}   \\
& \hspace{0.1cm} (b_{vs}^{\text{min}}-b_r) \cdot  \tau_{uvrs}  \leq  0, 			& u \in \mathcal{U}, v \in \mathcal{V},  r \in \mathcal{R}, s \in \mathcal{S} \label{eq:rate_resol_pairs_not_allowed_min_resol} \\
& \hspace{0.1cm} (b_r-b_{vs}^{\text{max}}) \cdot  \tau_{uvrs}  \leq  0, 		& u \in \mathcal{U}, v \in \mathcal{V}, r \in \mathcal{R}, s \in \mathcal{S} \label{eq:rate_resol_pairs_not_allowed_max_resol} \\
&\hspace{0.1cm} \sum_{u \in \mathcal{U}} \sum_{v \in \mathcal{V}}  \sum_{r \in \mathcal{R}} \sum_{s \in \mathcal{S}} b_r \cdot\tau_{udvrs} \leq C \cdot |\mathcal{U}|, \label{eq:tot_capac}  \\
&\hspace{0.1cm} \sum_{v \in \mathcal{V}}  \sum_{r \in \mathcal{R}} \sum_{s \in \mathcal{S}}  \beta_{vrs}  \leq K,  \label{eq:nr_repres}  \\
&\hspace{0.1cm} \sum_{u \in \mathcal{U}}  \gamma_{u} \geq  P \cdot |\mathcal{U}|,   \label{eq:satisUsers}  \\
&\hspace{0.1cm} \sum_{v \in \mathcal{V}} \sum_{r \in \mathcal{R}}  \sum_{s \in \mathcal{S}}   \tau_{uvrs} \geq  T_{\text{min}} \cdot \gamma_{u},   &\hspace{-1.2cm} u \in \mathcal{U} \label{eq:worst_satisfaction}  \\
&\hspace{0.1cm}\tau_{uvrs} \in [0,1], 								&\hspace{-3cm} u \in \mathcal{U}, v \in \mathcal{V},  r \in \mathcal{R},  s \in \mathcal{S}  \\
&\hspace{0.1cm} \alpha_{uvrs} \in \{0,1\}, 								&\hspace{-3cm} u \in \mathcal{U}, v \in \mathcal{V},  r \in \mathcal{R},  s \in \mathcal{S}  \\
&\hspace{0.1cm}  \beta_{vrs} \in \{0,1\}, 									&  v \in \mathcal{V},  r \in \mathcal{R},  s \in \mathcal{S} 
\end{align}
\end{subequations}
\end{ilp}

The objective function~\eqref{eq:objective} maximizes the sum of the user 
satisfactions averaged over time.  The constraints~\eqref{eq:alpha_on}, 
\eqref{eq:beta_off} and \eqref{eq:beta_on} set up a consistent relation between 
the decision variables $\tau$, $\alpha$ and $\beta$. The 
constraint~\eqref{eq:cdf} guarantees that a user $u$ plays a given 
representation only for the percentage of time during which the maximal user 
throughput is larger than the encoding rate $r$ of the representation. The 
constraint~\eqref{eq:demand} establishes that a user $u$ can play only those 
representations of the requested video $v_u$ with spatial resolutions compatible 
with the user display size $s_u$, that is  only representations with   
resolutions $\{s_u-1, s_u, s_u+1\}$ are allowed.  Namely, the possible 
down-sampling/up-sampling operations at the rendering are constrained to the  
resolutions that are immediately adjacent to the nominal user display size 
$s_u$. The constraints~\eqref{eq:rate_resol_pairs_not_allowed_min_resol}  and 
\eqref{eq:rate_resol_pairs_not_allowed_max_resol} force to zero some $\alpha$   
variables in order to ensure that each user $u$ only watches representations of 
video $v$ at resolution $s$ encoded at the bit rates in the range between the 
minimal and maximal admissible rates for the video $v$ and the resolution $s$. 
The constraint~\eqref{eq:tot_capac} guarantees that the sum of the average 
bit rates downloaded by all users  is lower than the overall CDN budget $C 
\cdot |\mathcal{U}|$. The constraint~\eqref{eq:nr_repres} fixes the maximal 
number of representations made available at the server. Finally, the 
constraints~\eqref{eq:satisUsers} and~\eqref{eq:worst_satisfaction} force the 
system to serve at least a certain percentage $P$ of users during a certain 
percentage  of time $T_{\text{min}}$. 

A simplified version of this model could be easily derived for those scenarios 
where the information of   client bandwidths is a priori available.  For instance, if the content provider obtains the 
access bandwidth of the end-users at one time instant  or estimates this 
bandwidth  by using a representative statistics, like an average value, a median 
value or a $n$th percentile. In these cases, the content provider uses a unique 
value $c_u$ to model the link capacity of each user $u$, motivating the 
introduction of the following changes in the ILP formulation of 
\eqref{formulation_one_resol_resolbased}: (i) the cumulative distribution 
function of each user $u$ is assumed to be a unit step function centered at the 
value $c_u$, (ii) the variables $\tau_{uvrs}$ are forced to be binary,  and 
(iii) $T_{\text{min}}$ is fixed to 1. We must note that the resulting 
formulation becomes  equivalent to the ILP introduced in \cite{Toni:C14}.


\section{Numerical Analysis Settings}
\label{sec:evaluation-settings}
We now describe the simulation framework that has been used to study  the 
\ac{ILP} introduced in Section~\ref{sec:ilp-model} for computing the optimal 
representations sets. We have used the generic solver IBM ILOG CPLEX~\cite{CPLEX} 
to solve different instances of the \ac{ILP} and to compare the optimized 
representations to the ones recommended by manufacturers and content providers. 
We have considered different \emph{configurations} in our study of the system 
performance. 
These scenarios are not meant to be an exhaustive list covering all possible 
cases.  Rather they   illustrate how the optimal set of representations changes in several  realistic  cases.

\begin{table}[t]
\centering
\begin{tabular}{ll}
\toprule
Video Type & Video Name\\
\midrule
Documentary &  Aspen, Snow Mountain  \\
Sport       &  Rush Field Cuts,Touchdown Pass,    \\
Cartoon     &  Big Buck Bunny, Sintel Trailer     \\
Movie       &  Old Town Cross      \\
\bottomrule
\end{tabular}
\caption{Test videos and corresponding types.}\label{tab:test_sequences}
\end{table} 

\begin{table}[t]
\centering
\begin{tabular}{cc}
\toprule
Resolution Name & Width x Height \\
\midrule
224p &  400x224  \\
360p &  640x360  \\
720p & 1280x720  \\
1080p& 1920x1080 \\ 
\bottomrule
\end{tabular}
\caption{Spatial resolutions used.}\label{tab:test_resolutions}
\end{table}

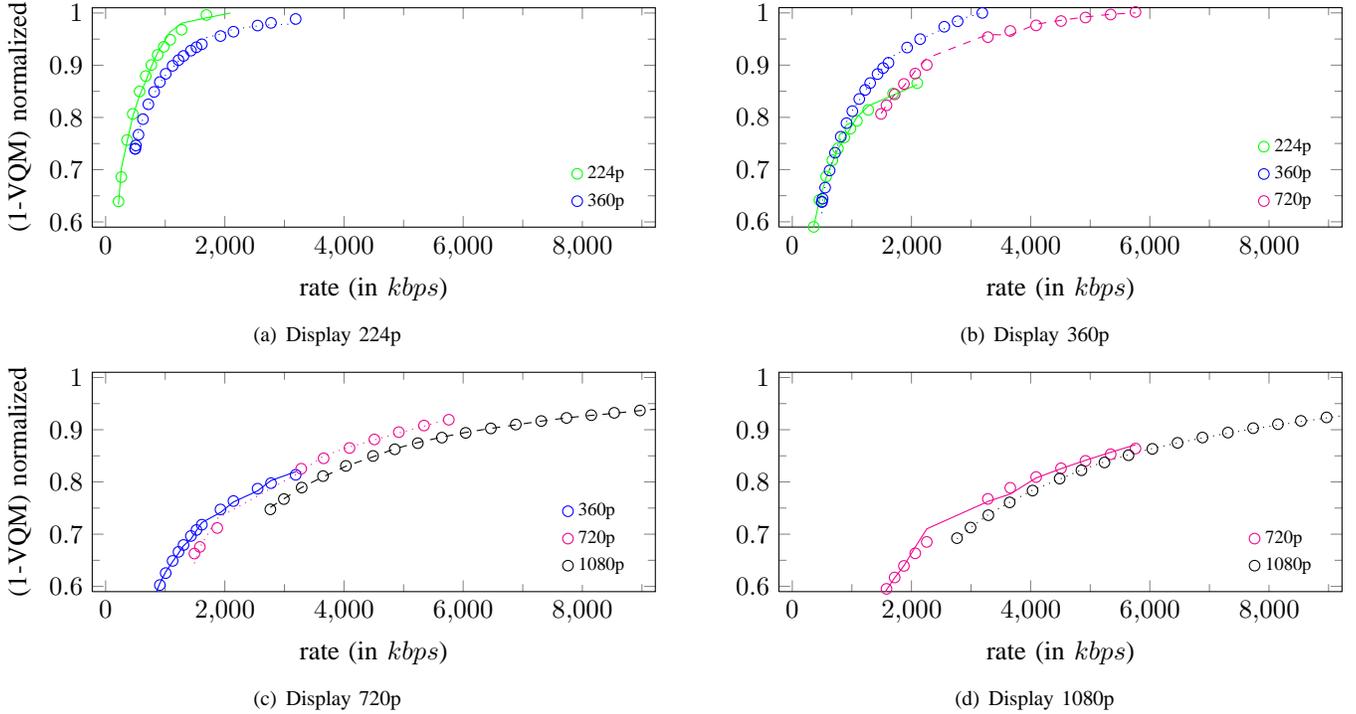
\begin{figure}
  \centering

  \newcommand\largeur{0.5}
\newcommand\hei{4.5cm}

\subfigure[Display 224p]{\label{qoe-sport_224}
    \begin{tikzpicture}[scale=1]
     \begin{axis}[
       width=\largeur\columnwidth,
       height=\hei,
        ymin=0.6,   
        ymax=1,  
        xmin=0,
        xmax=9000,
        ylabel near ticks, xlabel near ticks,
        minor tick num = 1,
        xlabel={rate (in $kbps$)},
        ylabel={(1-VQM) normalized},
        enlargelimits=0.025,
        mark options = {solid},
        legend style={font=\scriptsize,legend pos=south east, legend
          columns = 1, draw=none},
        legend cell align=left
]

        \pgfplotstableread{QoE_resol/sport_224_224.csv}\loadedtable;
        \addplot[no marks,green] table[x index=0, y index=1] {\loadedtable};
        \addplot[mark=o,only marks,green] table[x index=0, y index=2] {\loadedtable};

        \pgfplotstableread{QoE_resol/sport_224_360.csv}\loadedtable;
        \addplot[no marks,dotted,blue] table[x index=0, y index=1] {\loadedtable};
        \addplot[mark=o,only marks,blue] table[x index=0, y index=2] {\loadedtable};

        \legend{,224p,,360p,}

      \end{axis}
    \end{tikzpicture}
}
\qquad
\subfigure[Display 360p]{\label{qoe-sport_360}
    \begin{tikzpicture}[scale=1]
     \begin{axis}[
       width=\largeur\columnwidth,
       height=\hei,
        ymin=0.6,   
        ymax=1,  
        xmin=0,
        xmax=9000,
        ylabel near ticks, xlabel near ticks,
        minor tick num = 1,
        xlabel={rate (in $kbps$)},
        enlargelimits=0.025,
        mark options = {solid},
        legend style={font=\scriptsize,legend pos=south east, legend
          columns = 1, draw=none},
        legend cell align=left
]
     
        \pgfplotstableread{QoE_resol/sport_360_224.csv}\loadedtable;
        \addplot[no marks,green]table[x index=0, y index=1] {\loadedtable};
        \addplot[mark=o,only marks,green] table[x index=0, y index=2] {\loadedtable};

        \pgfplotstableread{QoE_resol/sport_360_360.csv}\loadedtable;
        \addplot[no marks,dotted,blue]table[x index=0, y index=1] {\loadedtable};
        \addplot[mark=o,only marks,blue]  table[x index=0, y index=2] {\loadedtable};

        \pgfplotstableread{QoE_resol/sport_360_720.csv}\loadedtable;
        \addplot[no marks,dashed,magenta]table[x index=0, y index=1] {\loadedtable};
        \addplot[mark=o,only marks,magenta]  table[x index=0, y index=2] {\loadedtable};

        \legend{,224p,,360p,,720p}

      \end{axis}
    \end{tikzpicture}
}
\subfigure[Display 720p]{\label{qoe-sport_720}
    \begin{tikzpicture}[scale=1]
     \begin{axis}[
       width=\largeur\columnwidth,
       height=\hei,
        ymin=0.6,   
        ymax=1,  
        xmin=0,
        xmax=9000,
        ylabel near ticks, xlabel near ticks,
        minor tick num = 1,
        xlabel={rate (in $kbps$)},
        ylabel={(1-VQM) normalized},
        enlargelimits=0.025,
        mark options = {solid},
        legend style={font=\scriptsize,legend pos=south east, legend
          columns = 1, draw=none},
        legend cell align=left
]

        \pgfplotstableread{QoE_resol/sport_720_360.csv}\loadedtable;
        \addplot[no marks,blue] table[x index=0, y index=1] {\loadedtable};
         \addplot[mark=o,only marks,blue] table[x index=0, y index=2] {\loadedtable};

        \pgfplotstableread{QoE_resol/sport_720_720.csv}\loadedtable;
        \addplot[no marks,dotted,magenta] table[x index=0, y index=1] {\loadedtable};
         \addplot[mark=o,only marks,magenta]table[x index=0, y index=2] {\loadedtable};

        \pgfplotstableread{QoE_resol/sport_720_1080.csv}\loadedtable;
        \addplot[no marks,dashed,black]table[x index=0, y index=1] {\loadedtable};
        \addplot[mark=o,only marks,black]  table[x index=0, y index=2] {\loadedtable};

        \legend{,360p,,720p,,1080p}

      \end{axis}
    \end{tikzpicture}
}
\qquad
\subfigure[Display 1080p]{\label{qoe-sport_1080}
    \begin{tikzpicture}[scale=1]
     \begin{axis}[
       width=\largeur\columnwidth,
       height=\hei,
        ymin=0.6,   
        ymax=1,  
        xmin=0,
        xmax=9000,
        ylabel near ticks, xlabel near ticks,
        minor tick num = 1,
        xlabel={rate (in $kbps$)},
        enlargelimits=0.025,
        mark options = {solid},
        legend style={font=\scriptsize,legend pos=south east, legend
          columns = 1, draw=none},
        legend cell align=left
]

        \pgfplotstableread{QoE_resol/sport_1080_720.csv}\loadedtable;
        \addplot[no marks,magenta]  table[x index=0, y index=1] {\loadedtable};
        \addplot[mark=o,only marks,magenta] table[x index=0, y index=2] {\loadedtable};

        \pgfplotstableread{QoE_resol/sport_1080_1080.csv}\loadedtable;
        \addplot[no marks,dotted,black]table[x index=0, y index=1] {\loadedtable};
        \addplot[mark=o,only marks,black]  table[x index=0, y index=2] {\loadedtable};

           \legend{,720p,,1080p}

      \end{axis}
    \end{tikzpicture}
}
\caption{Curve fitting for all the considered  display resolutions for sport video. Lines are
  experimental  measures taken from the video while circles represent the model.}
  \label{fig:Curve_Fitting_sport}
\end{figure}

\subsection{User Satisfaction Evaluation}
We characterize each video at a given spatial resolution by one 
\emph{satisfaction function} that depends on the display resolution and 
expresses the \ac{QoE} as a function of the encoding rate.  Several works 
have investigated how to model this behavior but a uniformly accepted model is 
still missing~\cite{Zhan_Ma:ArXiv12}. In our case, we model the satisfaction 
function as a \ac{VQM} score~\cite{VQM_software}, which is a full-reference 
metric that has higher correlation with human perception than other MSE-based 
metrics, as  shown in~\cite{Bess:C13}. For spatial down/up-sampling in the 
video player, we adopt the Avisynth Lanczos filter~\cite{duchon1979lanczos}, 
which has been already adopted in existing video 
players/tools~\cite{Lancoz,VLCplayer}.

We have evaluated the \ac{VQM} score for four different types of test sequences 
available at~\cite{videotest}. Each of these test sequences corresponds to a 
representative video type as given in Table~\ref{tab:test_sequences}. The tested 
sequences have been encoded at different rates and at the resolutions described 
in Table~\ref{tab:test_resolutions}. Since the \ac{VQM} score ranges from $0$ 
to $1$ for the best and the worst QoE, respectively, we  associate  the user 
satisfaction level with the $(1-\text{VQM})$ score. The empirical measures 
obtained from evaluating the aforementioned sequences are depicted as continuous 
lines in Fig.~\ref{fig:Curve_Fitting_sport} for the sport video.  For the 
sake of brevity, here we depict only the sport video curves.   From 
these figures we can  better understand the video classification.  Video 
categorization is aimed at providing a rough but yet accurate notion of motion 
level of the video content.  For example, most sport sequences have a 
higher-motion level than most  documentary sequences.   This can be observed 
from the satisfaction curves, which are steeper for documentary  sequences than 
for sport ones. We provide the full set of
satisfaction curves and fitting models in the Online Appendix. 

From these measures, we derived a satisfaction function by fitting a function of 
the following form:
\begin{equation}\label{eq:QoE} 
    f_{uvrs} =1 - \left( m_{uvs} + \frac{n_{uvs}}{b_r+o_{uvs}}\right).
\end{equation} 
It represents  the  satisfaction level $f_{uvrs}$ of user $u$ receiving  video 
$v$ encoded at rate $r$ and resolution $s$ and displayed at size $s_u$. 
Table~\ref{tab:test_QoE}, in the Online Appendix, gives the parameters $m_{uvs}, n_{uvs},$ and $o_{uvs}$  
used in the  curve fitting process for each video $v$ and resolution $s$ to be 
displayed at size $s_u$. We recall that the parameter $b_r$ is  the nominal value in $kbps$ of 
the rate $r$. Note that the expression in Eq.~\eqref{eq:QoE} has an explicit 
dependency only on the encoding rate, while other parameters (video content 
information, encoding resolution, spatial down/up-sampling) are implicitly taken 
into account into  the model parameters $m_{uvs}, n_{uvs},$ and $o_{uvs}$.  The 
satisfaction curves evaluated from  Eq.~\eqref{eq:QoE} are identified by circles 
in Fig.~\ref{fig:Curve_Fitting_sport}. 

Note also that other satisfactions functions could be considered. However, as 
mentioned above, to this day  there are no commonly accepted QoE 
metrics~\cite{Fra:J12,Zhan_Ma:ArXiv12}, and the metric considered in 
Eq.~\eqref{eq:QoE} takes into account all factors that are critical in our 
problem formulation.

\pgfplotscreateplotcyclelist{cl_users}{%
    thick,black,solid,every mark/.append style={solid,fill=black},mark=*\\%
    thick,black!60!white,dotted,every mark/.append style={solid,fill=black!60!white},mark=square*\\%
    thick,black!30!white,dashed,every mark/.append style={solid,fill=black!30!white},mark=triangle*\\%
} 

\noindent
\begin{figure}[t]
	\centering
	\subfigure{
	\begin{tikzpicture}
	\begin{axis}[ylabel=Data Received (MB),xlabel=Chunk Number,cycle list name=cl_users,
	ylabel near ticks,	xlabel near ticks,	label style={font=\scriptsize},
	tick label style={/pgf/number format/fixed,/pgf/number format/1000 sep = \thinspace,font=\scriptsize},
	scaled y ticks=false,height=4cm,width=4.6cm,]
	
	\addplot table[col sep=comma,y=client_received_mb,x=client_iteration]{simulations/dash-user-0.csv};
	
	\addplot table[col sep=comma,y=client_received_mb,x=client_iteration]{simulations/dash-user-1.csv};
	
	\addplot table[col sep=comma,y=client_received_mb,x=client_iteration]{simulations/dash-user-3.csv};
	\end{axis}
	\end{tikzpicture}}
	\subfigure{
	\begin{tikzpicture}
	\begin{axis}[ylabel=Download Time (s),xlabel=Chunk Number,cycle list name=cl_users,
	ylabel near ticks,	xlabel near ticks,	label style={font=\scriptsize},	scaled y ticks=false,	
	extra y ticks={2},	extra y tick labels={},	extra y tick style={grid=major}, legend columns=3, 
	legend style={font = \scriptsize,at={(0.5,1.4)},anchor=north,/tikz/every even column/.append style={column sep=0.5cm}}, 
	extra description/.code={\node[font=\tiny,color=gray] at (0.15,0.65) {expected};},
	tick label style={/pgf/number format/fixed,/pgf/number format/1000 sep = \thinspace,font=\scriptsize},
	height=4cm,width=4.6cm,]
	\addlegendentry{IP 1} 
	\addplot table[col sep=comma,y=client_elapsed,x=client_iteration]{simulations/dash-user-0.csv};
	\addlegendentry{IP 2} 
	\addplot table[col sep=comma,y=client_elapsed,x=client_iteration]{simulations/dash-user-1.csv};
	\addlegendentry{IP 3} 
	\addplot table[col sep=comma,y=client_elapsed,x=client_iteration]{simulations/dash-user-3.csv};
	\end{axis}
	\end{tikzpicture}}
	\subfigure{
	\begin{tikzpicture}
	\begin{axis}[ylabel=Download Rate (Mbps),xlabel=Chunk Number,cycle list name=cl_users,
	ylabel near ticks,xlabel near ticks,label style={font=\scriptsize},scaled y ticks=false,	
	tick label style={/pgf/number format/fixed,/pgf/number format/1000 sep = \thinspace,font=\scriptsize},
	height=4cm,width=4.6cm,]
	\addplot table[col sep=comma,y=bw_mbps,x=client_iteration]{simulations/dash-user-0.csv};
	\addplot table[col sep=comma,y=bw_mbps,x=client_iteration]{simulations/dash-user-1.csv};
	\addplot table[col sep=comma,y=bw_mbps,x=client_iteration]{simulations/dash-user-3.csv};
	\end{axis}
	\end{tikzpicture}}
      \caption{Example of three sessions from the dataset~\protect\cite{Basso:C14}.  The figure on the left represents the amount of data received for each chunk.  In the middle figure, the time to download each chunk is illustrated.  The right figure shows the download rate of users, calculated by the amount of data received (left) divided by the download time (middle). }
	\label{fig:dash-users}
\end{figure}
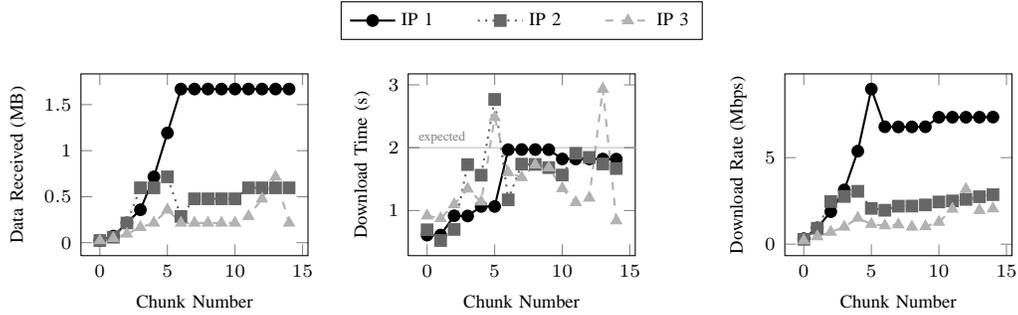

\subsection{User Population Characteristics} 
\label{sec:user-population}
A user $u \in \mathcal{U}$ is characterized by three parameters, which
we assign as follows:
\
\vspace{2mm}\newline
\noindent\textbf{The requested video stream $v_u$}. Users are randomly assigned 
to one of the four video types given in Table~\ref{tab:test_sequences}.  Each 
video type has the same probability ($1$ out of $4$) of being selected.
\
\vspace{2mm}\newline
\noindent\textbf{Statistical information about the streaming rate capacity 
$T_{ur}$}.  Recall that $T_{ur}$ is defined as the percentage of time a user 
$u$ has a streaming capacity greater than the encoding rate $b_r$. We model the 
streaming rate capacity with help of the dataset in~\cite{Basso:C14} that 
contains multiple measurements of thirty-second-long DASH 
sessions from thousands of geographically distributed IP addresses.  Each 
thirty-second measurement is associated with a user (IP address). Note that we 
have used users that have many thirty-second measurements in the dataset.  From 
the measurements in the dataset, we infer the \emph{download rate} of each IP 
address for every chunk of the session.  Fig.~\ref{fig:dash-users} illustrates 
the process we have used to compute the rate for each user. On the left, we show 
the number of bytes in every two-second-long chunk. In the middle, we show the 
time it takes for each IP address to entirely download the chunk. These two 
values are from the dataset. On the right, we compute the download rate for each 
IP address and for each chunk. Note that we do not consider the first five 
chunks of each session, so that the rump-up phase of each session is ignored. 
After we have computed the download rate of each user, we select a 
representative population of users from the $23,\!008$ distinct IP addresses 
from the dataset. We first filter the whole population and keep only the IP 
addresses of those clients whose  $75$th percentile of the download rate is 
lower or equal to $8\, Mbps$, which is the maximum encoding rate in our encoded 
set. Out of these selected users we then choose the IP addresses having the 
largest number of sessions. At the end, each user in our simulation framework  
is associated with one of the selected IP addresses. The streaming rate $T_{ur}$ 
of this user is computed from all the measurements in the dataset for the 
corresponding IP address.
\
\vspace{2mm}\newline
\noindent\textbf{The display size (spatial resolution) $s_u$}.  
We  categorize users into classes based on the display size  $s_u$ as 
follows. We assume that each user is characterized by the distribution of the 
available streaming rate, as described above, based on the dataset 
in~\cite{Basso:C14}. Based on their rate,  users request a display size  that fits the average available link capacity. For example, we associate users to  a display size of 224p if the $75\%$   of  the user link realizations can be found below $1,\!575\, kbps$. More formally, users 
with the $75$th percentile of download rate lower than $1,\!575\, kbps$ 
(respectively $2,\!400 \, kbps$, $4,\!500\, kbps$, and $6,\!750\, kbps$) are 
associated to a display size of 224p (respectively 360p, 
720p and 1080p). For the sake of clarity, we use the name of   few devices 
to indicate each class of users: smartphone for the users associated with the 
224p resolution, tablet for 360p, laptop for 720p and \ac{HDTV} for 1080p. Note that 
the device-based label  we assign to users is not a strict categorization, it 
is rather a shorthand to identify each group of users as each group is 
homogeneous in terms of requested resolution and link capacity statistics. 
This classification leads to a scenario with $90$ (respectively $67$, $161$, 
and $182$) users in the smartphone category (respectively tablet, laptop, and 
 \ac{HDTV}).

\subsection{Default ILP Settings}
We detail now the default settings used in the ILP instances studied in  the 
numerical analysis. These settings remain unchanged unless otherwise mentioned. 
First, the video catalog $\mathcal{V}$ and spatial resolution set $\mathcal{S}$ 
correspond to the video sequences and resolutions indicated in 
Table~\ref{tab:test_sequences} and Table~\ref{tab:test_resolutions}. The 
satisfaction coefficients $f_{vrs}$ are fixed for each triple ($v$, $r$, $s$) 
according to the satisfaction curves extrapolated from Eq.~\eqref{eq:QoE} with 
the parameters defined in Table~\ref{tab:test_QoE}.

The set of encoding rates $  \mathcal{R}$ is computed based on the user 
satisfaction curves. In particular, for each video  $v$ at  resolution $s$ 
displayed at  a display size such that  $s=s_u$,  we identify as minimum and 
maximum encoding rates those achieving a user satisfaction of 
$0.6$ and $1$, respectively. The range $[0.6-1]$ is then discretized with  a 
uniform step. In  our case,  a step of $0.025$ is considered, for a total of 
$17$ discrete values of the satisfaction function. For each of these satisfaction values, using 
these values in Eq.~\eqref{eq:QoE} with parameters in Table~\ref{tab:test_QoE}, 
we identify the corresponding rate $r$. The minimum and maximum encoding rates 
$b_{vs}^{\text{min}}$ and $b_{vs}^{\text{max}}$ for each video $v$ and 
resolution $s$ derived with this procedure are shown in 
Table~\ref{tab:minMaxRates}.

 In our tests, we use the user population $\mathcal{U}$ described in 
Section~\ref{sec:user-population}, whose cardinality is 
$|\mathcal{U}| = 500$ users.  Larger populations could also be considered. Note however that the optimal representations sets derived in this work are highly sensitive to the heterogeneity  of users' profiles, in terms of bandwidth and requested videos, and not necessarily to the population cardinality.  The average 
CDN budget capacity per user ($C$) is set to $1,\!000 \,Mbps$ unless otherwise specified. 
This  large value of $C$ implies that the system is not constrained by the 
overall CDN budget capacity. The maximum number of representations ($K$) is 
$132$, the fraction of users that must be served ($P$) is $0.90$ and the minimum 
fraction of time during which users should be served ($T_{\text{min}}$) is 
$0.20$.

Finally, we would like to the mention that, for instances created according to 
the above settings, CPLEX was able to solve the \ac{ILP} model in a few minutes 
on an Intel(R) Xeon(R) CPU E5640 @ 2.67GHz with 24 GB of RAM.

\subsection{Video Player   Controller}
\label{sec:subsec_simsteps}
Given a representations set (either the solution of the \ac{ILP} or one based 
on a specific recommendation), we need to evaluate the performance for 
realistic ``sessions'',  where a realistic video player  mimics the behavior 
of real video player implementing adaptive streaming technologies. To this end, 
we implement two different rate-adaptive controllers:
\begin{table*}
\centering
\scriptsize
\begin{tabular}{@{}c ccc@{} ccc@{} ccc@{} ccc @{}}
\toprule
& \multicolumn{2}{c}{224p} & \phantom{a} & 
  \multicolumn{2}{c}{360p} & \phantom{a} &
  \multicolumn{2}{c}{720p} & \phantom{a} &
  \multicolumn{2}{c}{1080p} \\
\cmidrule{2-3} \cmidrule{5-6} \cmidrule{8-9}
\cmidrule{11-12}
&  $b_{vs}^{\text{min}}$   & $b_{vs}^{\text{max}}$   &&
   $b_{vs}^{\text{min}}$  & $b_{vs}^{\text{max}}$   &&
   $b_{vs}^{\text{min}}$   & $b_{vs}^{\text{max}}$   &&
   $b_{vs}^{\text{min}}$  & $b_{vs}^{\text{max}}$   \\
\midrule
Movie       & 51   & 1961  && 67   & 2973 && 832 & 9378 && 1888 & 24803   \\
Sport       & 183   & 1766  &&  429 & 3190   && 1106   &11517 && 1976 &19471  & \\
Documentary & 116    &1488   && 231  & 2861  && 523 & 10607  && 1022    & 10945   \\
Cartoon     & 52    &1418 && 64   & 2006   && 451 & 5321 && 835  & 13133 \\
\bottomrule
\end{tabular}
\caption{Minimum and maximum encoding rates in $kbps$.}\label{tab:minMaxRates}
\end{table*}
\ \vspace{2mm}
\newline
{\bf ILP controller.}
 Among the representations available at
the server, each user asks for the one with the highest level
function among the ones with the encoding rate lower than or equal to the user
capacity. If no representation is available, the user is not served
(user in \emph{outage}) and the user satisfaction is set to zero. This
controller mimics the behavior that is considered in the \ac{ILP} formulation.
\ \vspace{2mm}
\newline
{\bf No-outage controller.}
As above, each client asks for the 
representation with the highest satisfaction level but with an encoding rate 
lower than or equal to the user link.  However, if no representation is 
available, the client asks for the representation that minimizes the excess 
between the requested encoding rate and the available bandwidth. This is 
justified by the fact that, in adaptive streaming scenarios, players are usually 
equipped with buffers that can temporally absorb small delays due to bandwidth 
fluctuations. We then assign to the user the satisfaction achieved by the 
requested representation, but we also keep track of the difference between the 
available bandwidth and the encoding rate selected by the client.  Note that in 
this second controller, each user is expected to be served, so that no outage 
is experienced.

\begin{table}
\centering
\scriptsize
\begin{tabular}{cccccccccccc}
\toprule
 & Representation  &   1  & 2 & 3  & 4  & 5  & 6 & 7  & 8  & 9 & 10 \\
\midrule
\multirow{2}{*}{Apple} & Rate ($kbps$) & $150$ & $200$ & $400$ & $600$ & $1,\!200$ & $1,\!800$ & $2,\!500$ & $4,\!500$ & $4,\!500$ & $6,\!500$ \\
& Resolution  & 224p & 224p & 224p & 360p & 360p & 720p & 720p & 720p & 1080p & 1080p \\  
\midrule
\multirow{2}{*}{Microsoft} & Rate ($kbps$) & $350$ & $400$ & $900$ & $1,\!250$ & $1,\!400$ & $2,\!100$ & $3,\!000$ & $3,\!450$ & $5,\!000$ & $6,\!000$ \\
& Resolution  & 224p & 224p & 224p & 360p & 720p & 720p & 720p & 720p & 1080p & 1080p \\  
\bottomrule
\end{tabular}
\caption{Representations 
 recommended by Apple and 
Microsoft.}\label{appleMicrosoftRepresentations}
\end{table}

 \begin{table*}
\centering
\scriptsize
\begin{tabular}{r ccccccccccc}
\toprule
Representation  &   1  & 2 & 3  & 4  & 5  & 6 & 7  & 8  & 9 & 10 & 11 \\
\midrule
Rate ($kbps$) & $150$ & $250$  & $350$ & $500$ & $650$ & $750$ & $1,\!000$  & 
$1,\!400$   & $1,\!500$ & $1,\!600$ & $1,\!750$ \\
Resolution  & 224p & 224p  & 224p  & 224p  &  224p & 224p      
&224p & 224p &224p & 224p & 224p \\  
\midrule
Representation & 12 & 13 & 14 & 15 & 16 & 17 & 18  & 19 & 20 & 21 & 22 \\
\midrule
Rate ($kbps$) & $250$ & $350$ & $500$ & $650$  & $750$  & $1,\!000$  & 
$1,\!400$   & $1,\!500$   & $1,\!600$   & $1,\!750$  & $1,\!000$ \\
Resolution & 360p & 360p  & 360p  & 360p  & 360p & 360p & 360p    
   & 360p &360p & 360p & 720p \\
\midrule
Representation & 23 & 24  & 25  & 26  & 27  &  28  & 29  & 30 & 31 & 32 & 33 \\ 
\midrule
Rate ($kbps$) & $1,\!400$  & $1,\!500$  & $1,\!600$ & $1,\!750$ &$2,\!350$  & 
$3,\!600$   & $1,\!500$  & $1,\!600$ & $1,\!750$  & $2,\!350$  & $3,\!600$ \\
Resolution  & 720p & 720p & 720p & 720p  & 720p & 720p &1080p     
  &1080p &1080p & 1080p & 1080p \\
\bottomrule
\end{tabular}
\caption{Representations 
recommended by 
Netflix.}\label{netFlixRepresentations}
\end{table*}

\section{how good are the recommended sets?}
\label{sec:numerical-analysis}
Today's system engineers generally select encoding parameters for the 
representations following  recommendations given by systems manufactures or 
content providers. These are typically  versatile enough to apply to any 
possible scenario but not fully optimized with respect to content or context 
information. In this section we provide results of a numerical analysis that 
addresses the following question: \emph{how good are the recommended 
sets?}

We focus on three recommended representations sets: 
Apple~\cite{apple-setting,apple-setting_1} for HTTP Live Streaming (HLS), 
Microsoft~\cite{micros-setting} for Smooth Streaming (see 
Table~\ref{appleMicrosoftRepresentations}), and 
Netflix~\cite{DBLP:conf/infocom/AdhikariGHVHSZ12,netFlix-setting} (see 
Table~\ref{netFlixRepresentations}).  Overall Microsoft and Apple recommend  
$10$ representations per video type, for a total of $40$ representations to be 
available at the server while Netflix recommends $33$ per video type, $132$ 
representations in total.  
Recommendations are compared with the optimized representations sets, namely the  
 solution of the ILP formulation in Eq. 
\eqref{formulation_one_resol_resolbased}. Optimal sets are evaluated  for  
different values of both the number of representations (parameter $K$) and   
the CDN budget (parameter $C$) in the ILP formulation. Recall that $C$ is 
the CDN budget \emph{per user}. Both the optimized and recommended 
sets are tested in the scenario described in Section~\ref{sec:evaluation-settings} 
for the ILP and the no-outage controllers. 


\subsection{ILP Controller}
\begin{figure}[t]
\begin{center}
\includegraphics[width=1\linewidth,draft=false]{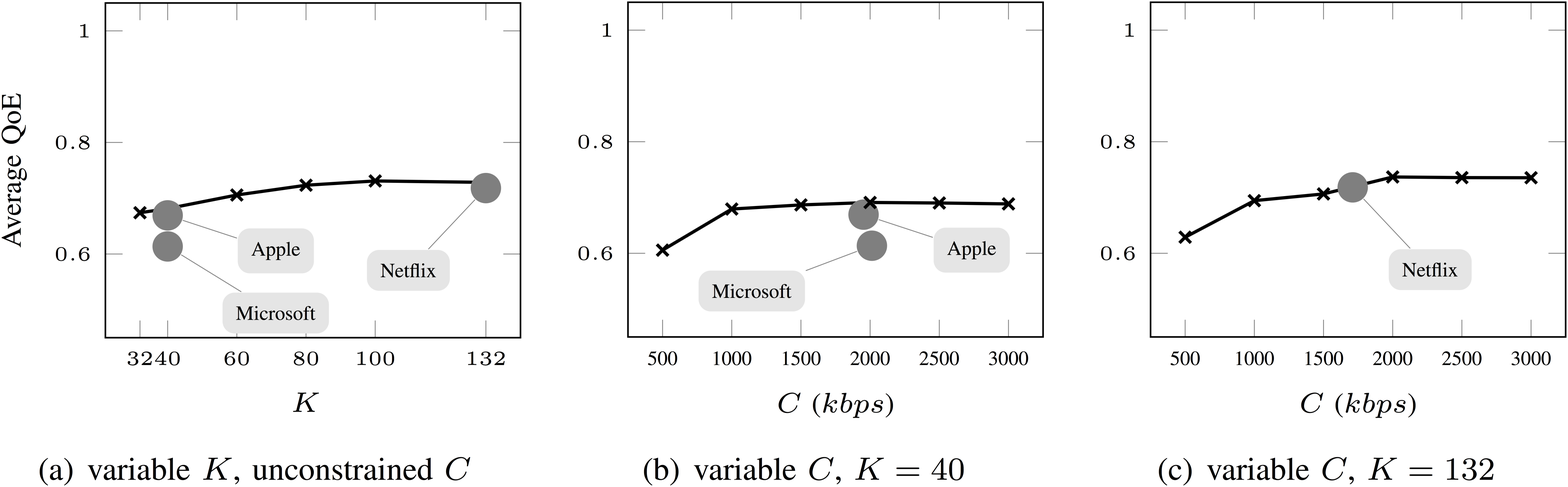}
\caption{Average QoE for various size of the representations set (a) and various CDN budget constraints (b and c). Since the   sets recommended by Apple, Netflix, and Microsoft  have fixed parameters, their    performances are only  given by a dot.} \label{fig:chunk-outage-ratio}
\end{center}
\end{figure}

The main performance metric that we consider in the ILP formulation is the 
average QoE per user, i.e., the average satisfaction. In 
Fig.~\ref{fig:chunk-outage-ratio},  we show the average QoE for various numbers of 
representations and various values for the  CDN budget. Our results show that 
the recommended sets perform reasonably well in terms of QoE, confirming what 
has been presented in~\cite{Toni:C14}. Typically, Apple and Netflix recommended 
sets are almost as good as the optimal one if the number of representations is 
not constrained.  However the optimized  sets can perform equally well with a 
smaller number of representations. Namely, Apple performance can be obtained 
with $K=32$ representations and Netflix ones with $K=80$. It is also worth 
noting that there exist representations sets that can also perform at least as 
well as Apple (respectively Microsoft) recommended sets with two (respectively 
four) times less overall bandwidth consumption (CDN budget).

 \begin{figure}[t]
\begin{center}
\includegraphics[width=1\linewidth,draft=false]{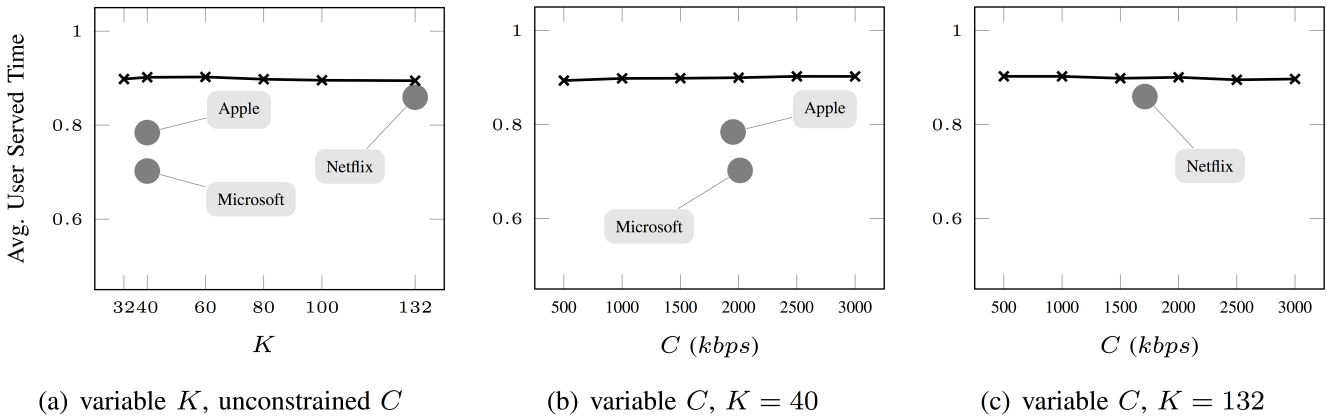}
 \caption{Average serving time per user (ratio of the number of
          downloaded chunks to the total number of chunks) for various numbers of representations $K$ (a) and  for various CDN budget  value  $C$ (b and c).} \label{fig:fairness_tot}
\end{center}
\end{figure} 

 \begin{table*}
\centering
\scriptsize
\begin{tabular}{rr ccccccccccccc }
\toprule
\multicolumn{13}{c}{Cartoon Type}  \\
\midrule
  Representation  &   1  & 2 & 3  & 4  & 5  & 6 & 7  & 8  &        \\
\midrule
  Rate ($kbps$) &  $52$ &       $82$     &   $283$     &   $451$    & $1,\!625$  &  $3,\!002$ &   $5,\!320$  & $834$    \\
 Resolution  &                224p  &  360p &    360p  &    720p &  720p     & 720p     &  720p & 1080p  \\  
\toprule
\multicolumn{13}{c}{Documentary Type}  \\
\midrule
  Representation &  1  & 2 & 3  & 4  & 5  & 6 & 7  & 8  &   9  \\
\midrule
   Rate ($kbps$) &   $115$      &   $187$     &   $230$   &   $415$    & $865$      &   $522$   &  $1,\!665$     & $2,\!838$  & $4,\!454$  \\
  Resolution &  224p  &   224p   &   360p &   360p  &  360p   &  720p  &   720p  &     720p  &    720p   \\
\toprule
\multicolumn{13}{c}{Movie Type}  \\
\midrule
   Representation &  1  & 2 & 3  & 4  & 5  & 6 & 7  & 8  &   9  & 10  & 11 &    \\
\midrule
   Rate ($kbps$) &  $51$   &   $178$    & $746$   &    $67$    &  $368$  &  $556$  &  $832$    &  $1,\!103$ & $1,\!596$ &   $2,\!424 $  & $3,\!645$	\\ 
  Resolution &  224p  &   224p &  224p &   360p  &   360p  &   360p  &  720p & 720p  &  720p  &  720p   & 720p          \\ 
\toprule
\multicolumn{13}{c}{Sport Type}  \\
\midrule
 Representation &  1  & 2 & 3  & 4  & 5  & 6 & 7  & 8  &   9   &  10  & 11 & 12\\
\midrule
   Rate ($kbps$) &  $183 $ &    $429$ & $ 771$ &   $955$  &    $1,\!371$ &  $1,\!824$  & $1,\!106$   &    $1,\!993$  & $2,\!876$  &    $3,\!755$  & $5,\!068$ &$ 7,\!239 $	\\ 
  Resolution &     224p &    360p &    360p &       360p &       360p  &     360p  &  720p     &   720p   & 720p &         720p &   720p &   720p    \\ 
\bottomrule
\end{tabular}
\caption{Representations 
 optimized for $K=40$ and $C=3$ Mbps.}\label{tab:opt_rep_K40_C3}
\end{table*}

To give a better understanding of the optimal representations set and how 
this differs from the recommended ones, Table~\ref{tab:opt_rep_K40_C3} shows the 
optimal solution (i.e., optimal representations set) of the \ac{ILP} for $K=40$ and 
$C=3\,Mbps$. By comparing the optimized set with the recommended ones 
(Table~\ref{appleMicrosoftRepresentations} and 
Table~\ref{netFlixRepresentations}) we can notice that $(i)$ the former does 
not have an equal number of representations per video type; $(ii)$  the 
encoding range changes according to  the video type. The different videos have different content characteristics 
(e.g., different motion levels) and they are better represented by a 
non-uniform allocation of the rate over a given encoding range. In the following 
section, we study in more details the behavior of the optimized representation 
sets.

In addition to the average QoE, we are interested in reducing the outage 
experienced by users. Fig.~\ref{fig:fairness_tot} shows the average serving 
time, i.e., the average time during which a user is served.  This serving time 
is normalized by the time duration of the session. Note that each users is 
served if it is able to request a representation at an encoding rate lower than 
its own available bandwidth. An average service time of $1$ means no outage, in 
other words, for each user there is always a representation that can be 
downloaded, i.e., there is always a representation encoded at a rate lower than 
the link capacity of that user.   Intuitively, since the ILP takes into account 
all the possible link capacities experienced by users over time,  it tends to 
offer a representations set well suited to channel dynamics. The results shown in 
Fig.~\ref{fig:fairness_tot} confirm this intuition. For every number of 
representations $K$    and every value of the  CDN 
budget $C$, the ILP can 
determine a representations set that covers well the range of user capacities. It 
is worth noting that, in terms of serving rate, the optimized set outperforms  
the representations sets recommended by Apple, Netflix, and Microsoft. This 
means that the ILP formulation, which takes into account the channel dynamics, 
provides a representations set more robust over time than recommended sets, 
leading to an average serving time of about $0.9$.

By combining the results of Fig.~\ref{fig:chunk-outage-ratio} and 
Fig.~\ref{fig:fairness_tot}, we conclude that the recommended sets perform 
well in terms of average QoE, but at a price: $(i)$ a high number of 
representations, $(ii)$ a high CDN budget, and $(iii)$ a low tolerance to 
variable downloading rates. The above results, however, consider a simplistic 
controller in which a user cannot be served when the requested rate exceeds  
the available bandwidth, i.e., when the  encoding rate is greater than the link 
capacity. To provide a fair comparison with the existing recommendation sets, we 
analyze the performance of the proposed set  also for the no-outage controller.  
 
\subsection{No-Outage Controller}
We now analyze the optimized representations set when the video players 
implement the no-outage controller.  This controller is probably closer to the
typical behavior of real clients than the ILP controller, but it does not 
exactly correspond to the model considered in the optimization problem in  
Eq.~\eqref{formulation_one_resol_resolbased}. We analyze in this case the 
\emph{\excess}, which is experienced any time a user requests a representation 
at rate $r$ that overshoots the channel link (i.e., $c\leq r$). The \excess \  
metric measures how much  the link capacity is overshoot and it is evaluated as $\max 
\left (0,\frac{r-c}{r}\right )$.  Note that we consider only the case in which 
the representation overshoots the channel bandwidth and not viceversa,   so we 
do not take into account negative values of the  \excess \ metric. Ideally, we 
would like to constantly experience a null  \excess, i.e., we always would like 
the requested representation to be supported by the channel link. In more 
realistic settings, in which the   \excess \ is not null, we would like it to 
be as low as possible. Indeed, a small  \excess \ can be easily absorbed  by the 
buffer that is  usually available at the client's player. On the contrary, a 
high  \excess \  might provoke video freeze for re-buffering.
  
 \begin{figure}[t]
\begin{center}
\includegraphics[width=1\linewidth,draft=false]{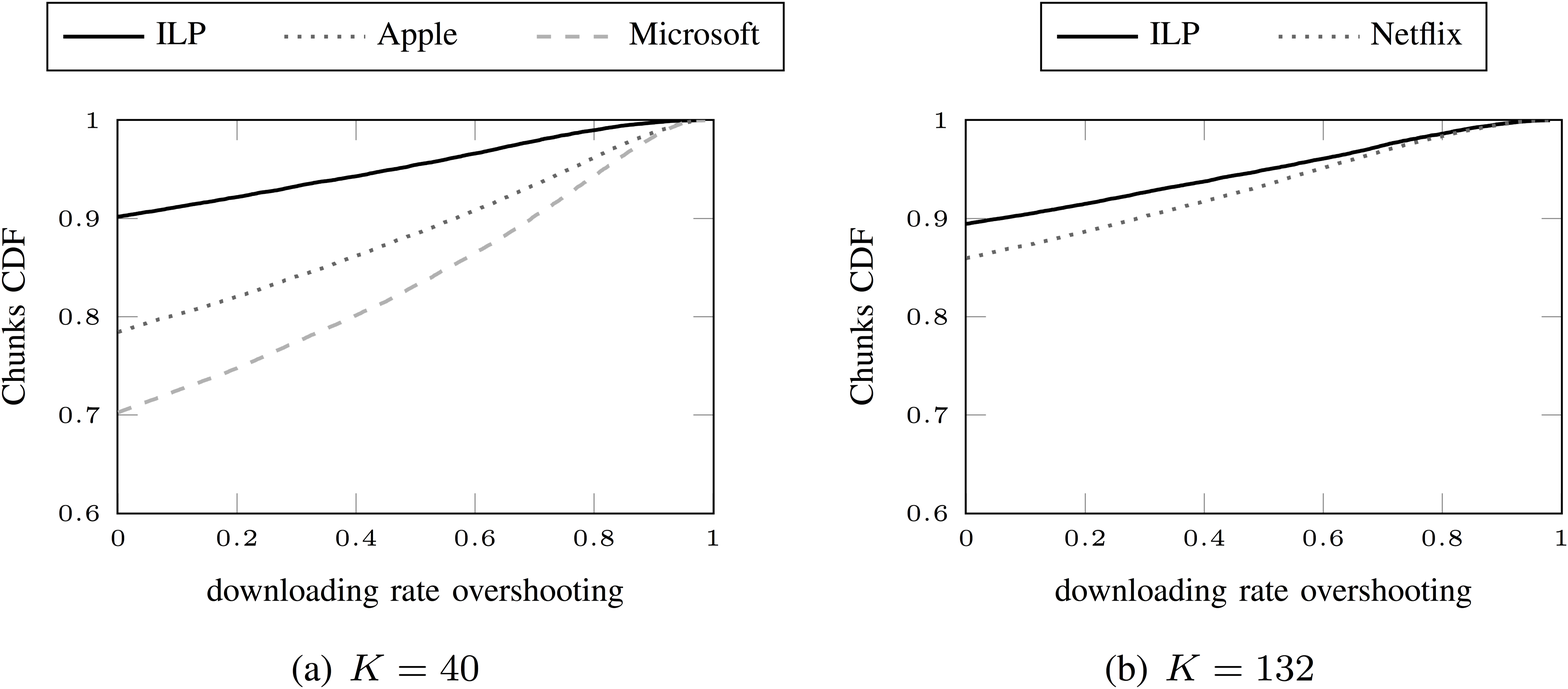}
 \caption{CDF of the total number of chunks vs. the  \excess. The number of representations of the ILP is
    equal to the number of representations for both Microsoft and
    Apple (a) and  for Netflix (b)  recommended sets. 
    }
    \label{fig:ratio_negativity}
\end{center}
 \end{figure}

Fig.~\ref{fig:ratio_negativity} shows the \ac{CDF} of the event of 
\excess \ for both default numbers of representations $K=40$ and $K=132$, which 
are used by the recommendations under consideration in this paper. Note that in 
the ILP formulation, the CDN budget is not constrained. In all cases, the 
recommended sets perform badly in comparison to what is obtained by the optimal 
representations set (although reducing the  \excess \ is \emph{not} the objective 
of the ILP formulation).   In more details, when $K=40$,  the ILP finds representations sets 
with $90\%$ of chunks having a null \excess, while it is less than $80\%$ for 
Apple and $70\%$ for Microsoft. 
Furthermore, a high number of chunks are downloaded with a high  \excess. For 
example, a  \excess \ of 0.5  means  that the video player needs a one-second 
buffer to compensate a two-second video downloading. For both Microsoft and 
Apple recommended sets, more than $15\%$ of chunks lead to such an annoying  
event. This result demonstrates that the optimal representations set   takes 
into account the channel variations without sacrificing the overall QoE.
 

\section{Guidelines}
\label{sec:guidelines}

In this section, we perform a comprehensive set of experiments with the 
objective of providing some \emph{guidelines} for selecting the representation 
sets. To obtain general guidelines, we need to consider multiple parameters in 
the users' population  and CDN characteristics in order to identify the main 
trends. This flexibility cannot be achieved with the population  described in 
Section.~\ref{sec:evaluation-settings} since it is extracted from a specific data 
set, corresponding to only one population. We thus use the same idea as 
in~\cite{Toni:C14}, which is to generate a synthetic user population 
characterized by a certain number of parameters. In this way we can explore 
different scenarios in a systematic manner by changing the values of these 
parameters consistently. In the following, we first describe how we generate the 
synthetic \emph{user populations sets}. Then, we describe the guidelines.


\subsection{Synthetic User Population Generation}
\label{sec:guidlines-user-population}

\begin{table}[t]
\centering
\scriptsize
\begin{tabular}{l ccc}
\toprule
Network & Minimum   & Maximum    & Attachment\\
Type    & Bandwidth & Bandwidth  & Probability\\
        & (in $Mbps$) & (in $Mbps$)  &\\
\midrule
WiFi (high load)       & 0.15  &   0.8     & 0.3 \\
3G          & 0.4   &   4       & 0.2 \\
ADSL-slow / WiFi (normal load)   & 0.3   &   3       & 0.1 \\
ADSL-fast   & 0.7   &   10      & 0.3 \\
FTTH        & 1.5   &   25      & 0.1 \\
\bottomrule
\end{tabular}
\caption{Different network types and corresponding 
parameters.}\label{tab:techno}
\end{table}

A user $u \in \mathcal{U}$ is characterized by three parameters:  requested 
video stream $v_u$, requested resolution $s_u$ and local network 
capacity $c_u$. These three parameters are assigned as follows:

\begin{itemize}
\item  $v_u$: users are randomly assigned to one of the four video types given  in Table~\ref{tab:test_sequences}.  Each video type has the same probability 
($1$ out of $4$) of being selected.
\item $s_u$: users are randomly assigned to one of four device types: 
smartphone, tablet, laptop and \ac{HDTV}.  Each device is associated to a  
resolution: 224p, 360p, 720p and 1080p for smartphone, tablet, laptop and 
\ac{HDTV}, respectively.  Again, each device type has the same probability 
($1$ out of $4$).
\item $c_u$: users are randomly assigned to one of the five network types in 
Table~\ref{tab:techno}, using the probability given in the last column of the table.  Once a 
user is associated to a given type of network, $c_u$ is selected as a uniformly 
distributed random value between minimum and maximum capacity (second and third 
column in Table~\ref{tab:techno}).
\end{itemize}

In comparison with the population described in 
Section.~\ref{sec:evaluation-settings} and simulated in 
Section.~\ref{sec:numerical-analysis}, the link capacity of a user $u$ is not 
characterized by a cumulative probability distribution. It rather assumes a 
constant value $c_u$.  The rationale behind that is to avoid generating complex 
populations with arbitrary channel variations, which challenges our original 
objective of having a common framework where population parameters can be 
easily modified. Thus, we run in the following the simplified version of the 
system model that is actually equivalent to the ILP in~\cite{Toni:C14}, and 
introduced in Section.~\ref{sec:ilp-model}.

\subsection{Results}
Studying the optimal representations sets evaluated across  different 
populations, we derive four \emph{guidelines}.

\vspace{0.35cm}
{\bf Guideline 1:  How many representation do we allocate per video type?}
     \emph{ The number of representations per video type 
    should be content-aware: a larger number of representations needs to be 
    dedicated to more complex video sequences.}
\vspace{0.15cm}

 \begin{figure}[t]
  \centering
  \subfigure[Resolution switching]{
\begin{tikzpicture}[scale=1]
	
	\begin{axis}[
		ytick pos=left,	
		scaled ticks=false,
		ylabel=nr of representations,
		xlabel=Resolution,
		ylabel near ticks,
		xlabel near ticks,
                ymin = 0,
		ymax=15,
		ybar=2.5pt,
		bar width=5pt,
                xtick={0,1,2,3},
		xticklabels={224p,360p,720p,1080p},
		enlarge x limits=0.2,
		enlarge y limits=0,
		height=0.3\textwidth,
		width=0.42\textwidth,
                legend style={font=\scriptsize,at={(1.05, 1.02)}, anchor=south
                  east, legend
                  columns = 4,draw=none},
                ymajorgrids=true,
                legend cell align=left,
		label style={font=\small}, 
                tick label style={font=\tiny},
                title
                style={at={(0.4,0)},color=gray!75!white,font=\large},
                area legend
	]	

        \pgfplotstableread{statReps/avgRepsPerResol_sc_neutral_resolSwitching_100_reps_3000_Kbps.txt }\loadedtable;

	\addplot[solid,fill=black] table[y index=1,x
        index=0]{\loadedtable};

	\addplot[solid,fill=gray!80] table[y index=2,x
        index=0]{\loadedtable};

	\addplot[solid,fill=gray!30] table[y index=3,x
        index=0]{\loadedtable};

	\addplot[solid,fill=gray!0] table[y index=4,x
        index=0]{\loadedtable};

	\end{axis}

\end{tikzpicture}
   \label{fig:column_switching}}
 %
%
  \centering
  \subfigure[No resolution switching]{
\begin{tikzpicture}[scale=1]
	
	\begin{axis}[
		ytick pos=left,	
		scaled ticks=false,
		ylabel=nr of representations,
		xlabel=Resolution,
		ylabel near ticks,
		xlabel near ticks,
                ymin = 0,
		ymax=10,
		ybar=2.5pt,
		bar width=5pt,
                xtick={0,1,2,3},
		xticklabels={224p,360p,720p,1080p},
		enlarge x limits=0.2,
		enlarge y limits=0,
		height=0.3\textwidth,
		width=0.42\textwidth,
                legend style={font=\scriptsize,at={(1.05, 1.02)}, anchor=south
                  east, legend
                  columns = 4,draw=none},
                ymajorgrids=true,
                legend cell align=left,
		label style={font=\small}, 
                tick label style={font=\tiny},
                title
                style={at={(0.4,0)},color=gray!75!white,font=\large},
                area legend
	]	

        \pgfplotstableread{statReps/avgRepsPerResol_sc_neutral_noResolSwitching_100_reps_3000_Kbps.txt}\loadedtable;

	\addplot[solid,fill=black] table[y index=1,x
        index=0]{\loadedtable};

	\addplot[solid,fill=gray!80] table[y index=2,x
        index=0]{\loadedtable};

	\addplot[solid,fill=gray!30] table[y index=3,x
        index=0]{\loadedtable};

	\addplot[solid,fill=gray!0] table[y index=4,x
        index=0]{\loadedtable};

       \legend{cartoon~~, documentary~~, sport~~, movie~~}
	\end{axis}

\end{tikzpicture}
   \label{fig:column_NO_switching}}
    \caption{ Number of representations per resolution, for each
    type of videos, for $K=100$.  } \label{fig:videoHist}
 \end{figure}
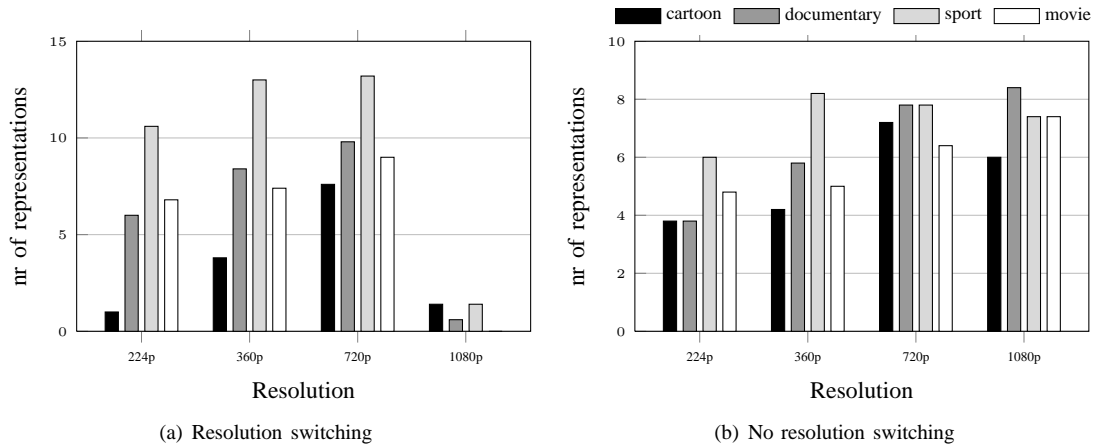 


A weakness of the recommended representations sets is that the number of 
available representations is the same for any video type, despite the different content characteristics.  Fig.~\ref{fig:videoHist} shows 
the  average number of representations dedicated to any video type as a function 
of the video resolution  for the optimal representations sets. Results are 
depicted  for two cases: $(i)$ when  users can play representations encoded at 
a resolution different from their display resolution (resolution switching), $(ii)$ when users are forced to only play videos encoded at their 
display resolution (no resolution switching). 

A first important observation is that using resolution switching can 
drastically affect the optimal representations set, as it can be observed by 
comparing Fig.~\ref{fig:column_switching} to 
Fig.~\ref{fig:column_NO_switching}. Given the user satisfaction curves in 
Fig.~\ref{fig:Curve_Fitting_sport}, when resolution switching is allowed 
(Fig.~\ref{fig:column_switching}), high resolution videos (i.e., 1080p) 
require higher rates yet the user satisfaction is almost identical to the one 
obtained using up-sampled  720p  videos, which require a lower rate. This 
means that, in our setting,   1080p  resolution videos are redundant in terms 
of quality offered to users, and when the total number of representations $K$ 
is limited, redundant representations are not included in the optimal set.  If 
no resolution switching is allowed (Fig.~\ref{fig:column_NO_switching}), 
videos at  1080p  are not redundant anymore, since they are the only resolution 
that can serve HD users. This comparison highlights the importance of taking 
into account a QoE metric in the optimization algorithm. A second consideration 
is that some videos clearly require more representations than 
others: about $38$ on average for sport videos while only about $13-14$  
representations on average for cartoon, for the case of switching 
resolution. This is justified  by the fact that sport videos have more 
complexity in the scene, leading to a  wider range of QoE values than for the 
cartoons.  Hence the need to have more representation for the sport video 
type rather than for the cartoon type. 

Rather than a uniform distribution of video types across users, we 
now study the optimal representations set for non-uniform popularity of video 
types. This should confirm  that these results are not biased by our default 
configuration.

Four video types are still considered, i.e., documentary, movie, sport and 
cartoon, but only  $10\%$ of  users watch the documentary,  another $10\%$ 
watch the movie, and the remaining watch either the cartoon of the sport 
video.  More precisely, $x$ is the percentage of users watching the sport 
video, and $0.8-x$ is the percentage of users watching the cartoon. In 
Fig.~\ref{fig:sport}, the parameter $x$ ranges from $0$ (no sport videos) to 
$0.8$ (no cartoon videos). We measure the distribution of the number of 
representations over the different videos when $K=100$.  In other words, 
Fig.~\ref{fig:sport} shows,  out of the $100$ representations, how many are  
dedicated to each type of video.  For example, when both sports and cartoon are 
requested by $40\%$ of the  population, representations are unequally 
distributed among videos ($39\%$ for  sport while $29\%$ for the cartoon). 
 Similar observations can be derived from Fig.~$9$, where we have considered the same video requests fraction as above ($0.1$ for documentary, $0.1$ for movie, $x$ for sport, and $0.8 - x$ for cartoon) but for the case in which resolution switching is allowed. Also in this case, when both sports and cartoon are requested by $40\%$ of the population, representations are unequally distributed among videos ($47\%$ for sport while $18\%$ for the cartoon).
Fig.~\ref{fig:sport} and Fig.~$9$(a)  confirm  our previous observations. Some videos, like 
cartoons, are under represented irrespective of their popularity. Cartoon 
videos are the $35.8\%$ of the total of representations even when they are 
requested by $60\%$ of the population.
This shows that the the content information, reflected in our case by the QoE 
user satisfaction function, is a critical input for selecting representation 
sets.

\begin{figure}[t]
%
\subfigure[]{
  \begin{tikzpicture}[scale=1]
    \begin{axis}[
      width= .5\columnwidth,
      height=5.5cm,
      ymin=0,   
      ymax=1,  
      xmin=0,
      xmax=0.8,
      minor tick num = 1,
      ylabel near ticks,
      xlabel={ratio of requests for sport videos   ($x$)  },
      ylabel={ratio of representations},
      xlabel style={font=\small},
      ylabel style={font=\small},
      xticklabel style={font=\small},
      yticklabel style={font=\small},
      enlargelimits=0,
      mark options = {solid},
      legend style={font=\scriptsize,at={(0.98, 1.02)}, anchor=south
        east, legend
        columns = 4,draw=none},
      legend cell align=left,
      extra y ticks={},
      extra y tick labels={},
      extra y tick style={grid=major},
      stack plots = y,
      area style,
      enlarge x limits=false
      ]

      \draw[thick,-|] (axis cs:0.4,0) -- node[fill=gray!80,font=\footnotesize,inner
      sep=1pt]{0.29}(axis cs:0.4,0.294);

      \draw[thick,-|] (axis cs:0.4,0.294) -- node[fill=gray!60,font=\footnotesize,inner
      sep=1pt]{0.15}(axis cs:0.4,0.4520);

\draw[thick,-|] (axis cs:0.4,0.4520) -- node[fill=gray!40,font=\footnotesize,inner
      sep=1pt]{0.16}(axis cs:0.4,0.61);

\draw[thick,-|] (axis cs:0.4,0.61) -- node[fill=gray!20,font=\footnotesize,inner
      sep=1pt]{0.39}(axis cs:0.4,1);      
      \pgfplotstableread{result/avgNbRepresPerVideo_noResolSwitchingsc_SportRate.txt}\loadedtable;      



      \addplot[fill=gray!80] table[x expr=\thisrowno{0}/100,y expr=\thisrowno{1}]{\loadedtable} \closedcycle;
      \addplot[fill=gray!60] table[x expr=\thisrowno{0}/100,y expr=\thisrowno{2}]{\loadedtable} \closedcycle;
      \addplot[fill=gray!40] table[x expr=\thisrowno{0}/100,y
      expr=\thisrowno{4}]{\loadedtable} \closedcycle;
      \addplot[fill=gray!20] table[x expr=\thisrowno{0}/100,y expr=\thisrowno{3}]{\loadedtable} \closedcycle;

      \legend{cartoon~~, documentary~~, movie~~, sport};

     \end{axis}
   \end{tikzpicture}
   \label{fig:sport}
 }
%
%
%
%
  \subfigure[]{
  \begin{tikzpicture}[scale=1]
    \begin{axis}[
      width= .5\columnwidth,
      height=5.5cm,
      ymin=0,   
      ymax=1,  
      xmin=0,
      xmax=0.8,
      minor tick num = 1,
      ylabel near ticks,
      xlabel={ratio of HDTV user in the population   ($y$) },
   ylabel={ratio of representations  },
      xlabel style={font=\small},
      ylabel style={font=\small},
      xticklabel style={font=\small},
      yticklabel style={font=\small},
      enlargelimits=0,
      mark options = {solid},
      legend style={font=\footnotesize,at={(0.95, 1.02)}, anchor=south
        east, legend
        columns = 4, draw=none},
      legend cell align=left,
      extra y ticks={},
      extra y tick labels={},
      extra y tick style={grid=major},
      stack plots = y,
      area style,
      enlarge x limits=false
      ]

      \draw[thick,-|] (axis cs:0.4,0) -- node[fill=gray!80,font=\footnotesize,inner
      sep=1pt]{0.25}(axis cs:0.4,0.25);

      \draw[thick,-|] (axis cs:0.4,0.25) -- node[fill=gray!60,font=\footnotesize,inner
      sep=1pt]{0.15}(axis cs:0.4,0.398);

      \draw[thick,-|] (axis cs:0.4,0.398) -- node[fill=gray!40,font=\footnotesize,inner
      sep=1pt]{0.22}(axis cs:0.4,0.622);

      \draw[thick,-|] (axis cs:0.4,0.622) -- node[fill=gray!20,font=\footnotesize,inner
      sep=1pt]{0.38}(axis cs:0.4,1);      
      \pgfplotstableread{result/avgNbRepresPerResol_noResolSwitchingsc_HDTVrate.txt}\loadedtable;
      
      \addplot[fill=gray!80] table[x expr=\thisrowno{0}/100,y expr=\thisrowno{1}]{\loadedtable} \closedcycle;
      \addplot[fill=gray!60] table[x expr=\thisrowno{0}/100,y expr=\thisrowno{2}]{\loadedtable} \closedcycle;
      \addplot[fill=gray!40] table[x expr=\thisrowno{0}/100,y expr=\thisrowno{3}]{\loadedtable} \closedcycle;
      \addplot[fill=gray!20] table[x expr=\thisrowno{0}/100,y expr=\thisrowno{4}]{\loadedtable} \closedcycle;
      \legend{224p~~~,360p~~~,720p~~~,1080p}
     \end{axis}
   \end{tikzpicture}
 \label{fig:hdtv}
 }
       \caption{Distribution of representations per video.  No resolution switching.}
 \end{figure}
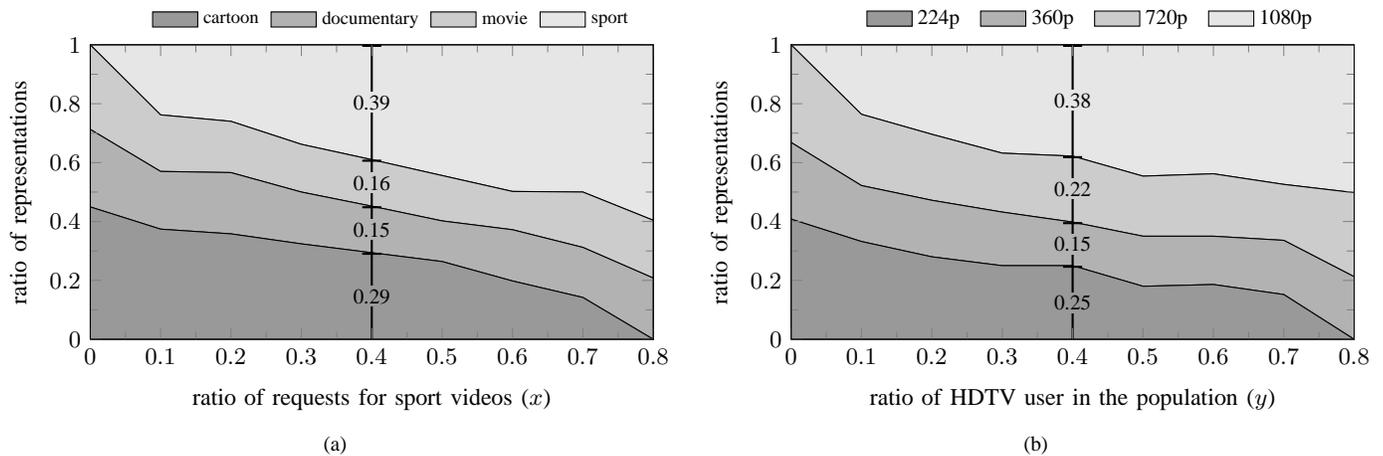

\begin{figure} 
\subfigure[]{
  \begin{tikzpicture}[scale=1]
    \begin{axis}[
      width= .5\columnwidth,
      height=5.5cm,
      ymin=0,   
      ymax=1,  
      xmin=0,
      xmax=0.8,
      minor tick num = 1,
      ylabel near ticks,
      xlabel={ratio of requests for sport videos   ($x$)  },
      ylabel={ratio of representations},
      xlabel style={font=\small},
      ylabel style={font=\small},
      xticklabel style={font=\small},
      yticklabel style={font=\small},
      enlargelimits=0,
      mark options = {solid},
      legend style={font=\scriptsize,at={(0.98, 1.02)}, anchor=south
        east, legend
        columns = 4,draw=none},
      legend cell align=left,
      extra y ticks={},
      extra y tick labels={},
      extra y tick style={grid=major},
      stack plots = y,
      area style,
      enlarge x limits=false
      ]

      \draw[thick,-|] (axis cs:0.4,0) -- node[fill=gray!80,font=\footnotesize,inner
      sep=1pt]{0.18}(axis cs:0.4,0.1866);

      \draw[thick,-|] (axis cs:0.4,0.1866) -- node[fill=gray!60,font=\footnotesize,inner
      sep=1pt]{0.17}(axis cs:0.4,0.3520);

\draw[thick,-|] (axis cs:0.4,0.3520) -- node[fill=gray!40,font=\footnotesize,inner
      sep=1pt]{0.16}(axis cs:0.4,0.5233);

\draw[thick,-|] (axis cs:0.4,0.5233) -- node[fill=gray!20,font=\footnotesize,inner
      sep=1pt]{0.47}(axis cs:0.4,1);      
      \pgfplotstableread{result/avgNbRepresPerVideo_resolSwitchingsc_SportRate.txt}\loadedtable;      %
      \addplot[fill=gray!80] table[x expr=\thisrowno{0}/100,y expr=\thisrowno{1}]{\loadedtable} \closedcycle;
      \addplot[fill=gray!60] table[x expr=\thisrowno{0}/100,y expr=\thisrowno{2}]{\loadedtable} \closedcycle;
      \addplot[fill=gray!40] table[x expr=\thisrowno{0}/100,y
      expr=\thisrowno{4}]{\loadedtable} \closedcycle;
      \addplot[fill=gray!20] table[x expr=\thisrowno{0}/100,y expr=\thisrowno{3}]{\loadedtable} \closedcycle;

      \legend{cartoon~~, documentary~~, movie~~, sport};

     \end{axis}
   \end{tikzpicture}
   \label{fig:sport2}
 }
%
%
%
%
  \subfigure[]{
  \begin{tikzpicture}[scale=1]
    \begin{axis}[
      width= .5\columnwidth,
      height=5.5cm,
      ymin=0,   
      ymax=1,  
      xmin=0,
      xmax=0.8,
      minor tick num = 1,
      ylabel near ticks,
      xlabel={ratio of HDTV user in the population   ($y$) },
   ylabel={ratio of representations  },
      xlabel style={font=\small},
      ylabel style={font=\small},
      xticklabel style={font=\small},
      yticklabel style={font=\small},
      enlargelimits=0,
      mark options = {solid},
      legend style={font=\footnotesize,at={(0.95, 1.02)}, anchor=south
        east, legend
        columns = 4, draw=none},
      legend cell align=left,
      extra y ticks={},
      extra y tick labels={},
      extra y tick style={grid=major},
      stack plots = y,
      area style,
      enlarge x limits=false
      ]
      
      \draw[thick,-|] (axis cs:0.4,0) -- node[fill=gray!80,font=\footnotesize,inner
      sep=1pt]{0.27}(axis cs:0.4,0.276667);

      \draw[thick,-|] (axis cs:0.4,0.276667) -- node[fill=gray!60,font=\footnotesize,inner
      sep=1pt]{0.20}(axis cs:0.4,0.4767);

      \draw[thick,-|] (axis cs:0.4, 0.4767) -- node[fill=gray!40,font=\footnotesize,inner
      sep=1pt]{0.48}(axis cs:0.4,0.956);

      \draw[thick,-|] (axis cs:0.4,0.956) -- node[fill=gray!20,font=\footnotesize,inner
      sep=1pt]{0.043}(axis cs:0.4,1);      
      \pgfplotstableread{result/avgNbRepresPerResol_resolSwitchingsc_HDTVrate.txt}\loadedtable;

      \addplot[fill=gray!80] table[x expr=\thisrowno{0}/100,y expr=\thisrowno{1}]{\loadedtable} \closedcycle;
      \addplot[fill=gray!60] table[x expr=\thisrowno{0}/100,y expr=\thisrowno{2}]{\loadedtable} \closedcycle;
      \addplot[fill=gray!40] table[x expr=\thisrowno{0}/100,y expr=\thisrowno{3}]{\loadedtable} \closedcycle;
      \addplot[fill=gray!20] table[x expr=\thisrowno{0}/100,y expr=\thisrowno{4}]{\loadedtable} \closedcycle;
      \legend{224p~~~,360p~~~,720p~~~,1080p}
     \end{axis}
   \end{tikzpicture}
 \label{fig:hdtv2}
 }
       \caption{Distribution of representations per video.  Resolution switching.}
 \end{figure}
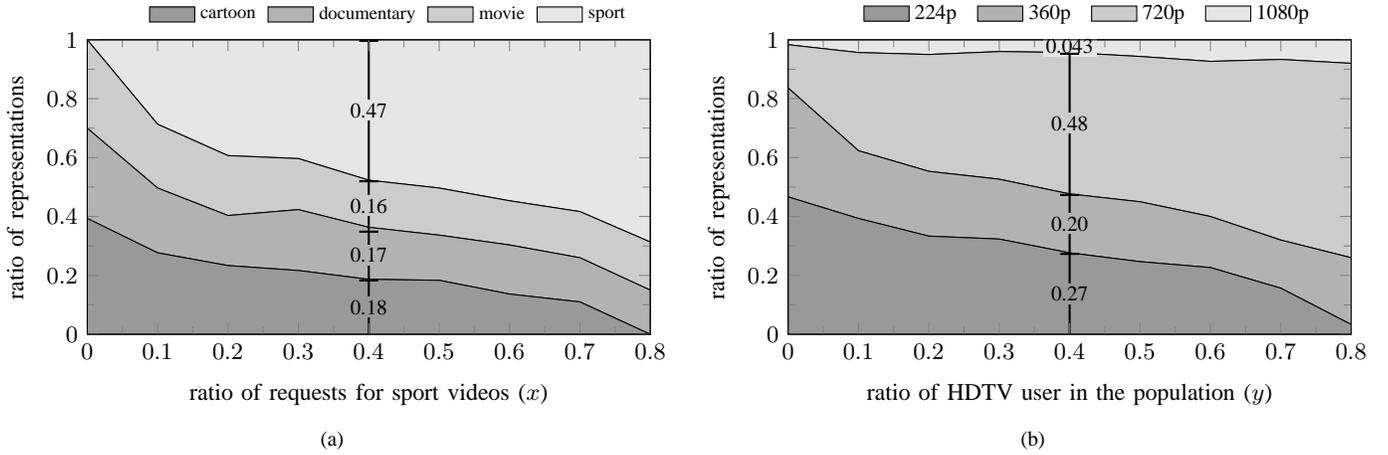


\vspace{0.35cm}
{\bf Guideline 2: For each video, how do we allocate the available representations across resolution?}     \emph{The distribution of the representation across resolutions should follow 
the distribution of user population across resolutions, putting an  
emphasis on the largest distributions.}
\vspace{0.15cm}

 \begin{figure}[t]
\begin{center}
\includegraphics[width=1\linewidth,draft=false]{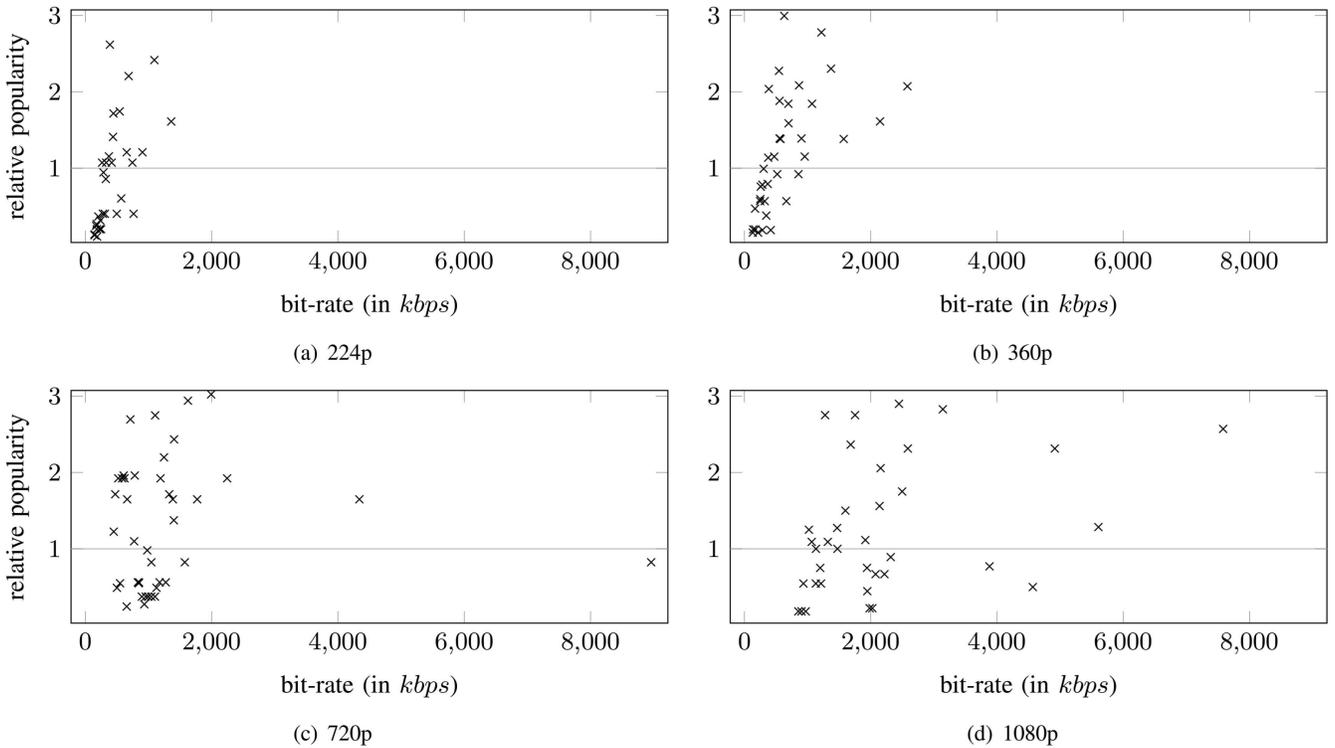}
\caption{Relative popularity of representations (number of users requesting 
a given representation with respect to the average number of users requesting 
any representation in the resolution of said representation) vs. bit rate. No resolution switching.}
   \label{fig:cloudpoint}
\end{center}
 \end{figure}

 \begin{figure}[t]
\begin{center}
\includegraphics[width=1\linewidth,draft=false]{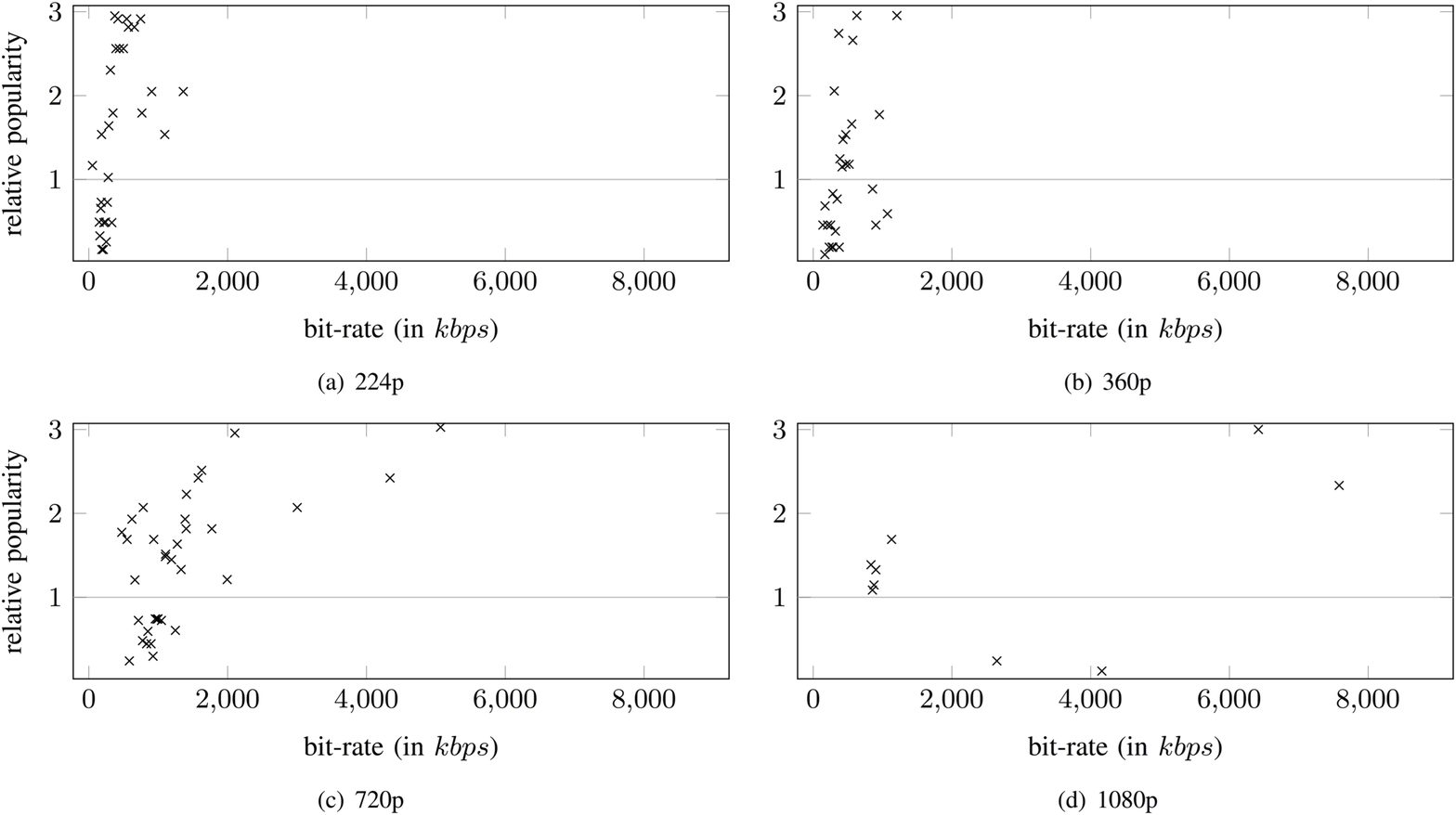}
    \caption{Relative popularity of representations (number of users requesting 
a given representation with respect to the average number of users requesting 
any representation in the resolution of said representation) vs. bit rate. Resolution switching}
   \label{fig:cloudpoint_switchi}
\end{center}
 \end{figure}

For a first analysis of the representation distribution per resolution, we can 
refer again to Fig.~\ref{fig:column_NO_switching}. For a given video, the 
number of representations increases with the resolution, but the increase is not 
substantial. Although the number of representations for sport videos is higher 
than for cartoon, we find here that there are on average $6$ representations at 
 224p  and $7.4$ at  1080p. This is however   not a major 
trend.

To dig deeper in the trend of representations distributions  across resolutions, we 
change the ratio of users' devices in the population. Similarly  to the 
 ratio of  users' videos in the population in 
Fig.~\ref{fig:sport},  we consider an unequal allocation of users to devices: 
$10\%$ of the population is identified as tablet users, $10\%$ as laptop users, 
and the remaining $80\%$ is shared between smartphone and HDTV users. We 
denote by $y$ the portion of HDTV users and $0.8-y$ portion of smartphone  
users.  Fig.~\ref{fig:hdtv} and   Fig.~$9$(b)
show  the ratio of representations for every  
resolution, for the no switching and switching case, respectively. The impact of the heterogeneity of users on the 
distribution of resolutions is less significant than for the popularity of 
videos.   The evolution of the ratio of representations per resolution follows 
the evolution of the distribution of devices in the user population. We also observe 
a slight over-representation of higher resolutions independently of  the ratio 
of HDTV users.

\vspace{0.35cm}
{\bf Guideline 3: For each video at a given resolution, how do we allocate the available representations across the encoding rates?} \emph{The higher is the resolution, the wider should be the range of encoding rates. Moreover, regardless the resolution, at least one  representation   encoded at lowest allowed rate    should  always be included. }
\vspace{0.15cm}

With the proposed \ac{ILP} formulation, we obtain an optimal set that maximizes 
the average user satisfaction. However, system engineers are also interested in 
maintaining consistency in their systems, trying to avoid for example that one 
representation is accessed by a lot of users although another representation 
serves only a few users. In Fig.~\ref{fig:cloudpoint}, not only we get some 
valuable insights about the range of bit rates in the optimal representation 
sets, but we can also analyze the ``popularity'' of each representation.

We define the \emph{relative popularity} as a value that indicates whether a 
representation is ``over-assigned'' (relative popularity greater than one) or 
``under-assigned'' (relative popularity lesser than one). In particular, let $L$ 
be a set of representations for a given video and a given resolution. Let $l$ be 
one representation in $L$. Let $n_L$ be the number of users who watch said 
video at said resolution. The average number of users per representation, 
which is hereafter noted by $n^{avg}_L$, is given by $\frac{n_L}{|L|}$. Let $n_l$ 
be the number of users assigned to representation $l\in L$. The relative 
popularity of the representation $l \in L$ is simply: 
\begin{equation*}
\frac{n_l}{n^{avg}_{L}}
\end{equation*}
In Fig.~\ref{fig:cloudpoint}, we gather the results of several realizations of 
the user population.  No resolution switching is allowed in this figure. We denote each realization as one run and we provide 
results for a total of five runs.  One mark shows that one representation has 
been created in one of the five runs for one of the videos. For each mark, we 
show the bit rate and the relative popularity of the representation.

 \begin{figure}[t]
\begin{center}
\includegraphics[width=1\linewidth,draft=false]{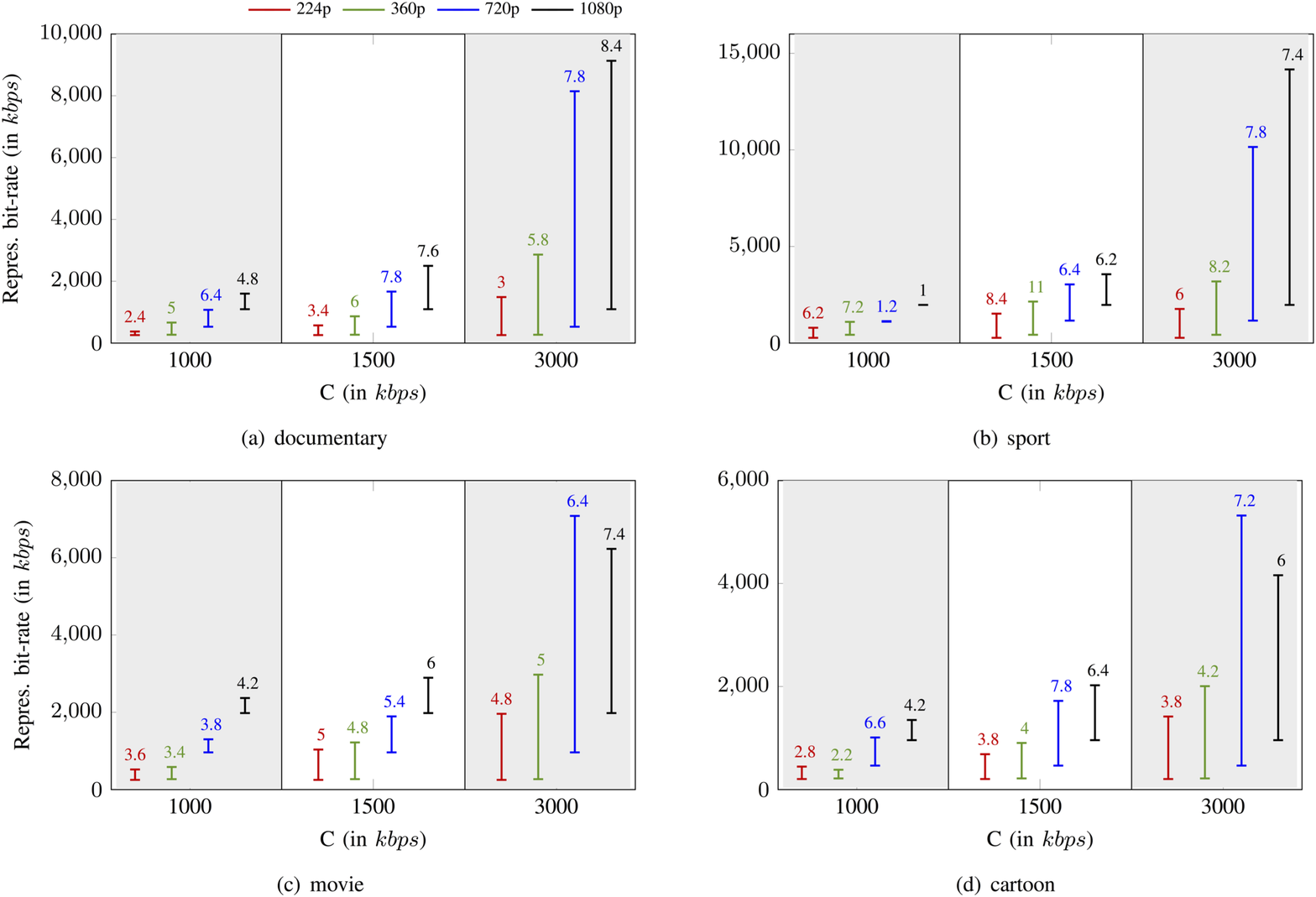}
    \caption{Range of representations bit-rate when $C$ is limited. 
 Bars are bounded, at the  bottom  (top), by the average minimum (maximum) value. The number over the bars indicates the   average number of representations for the resolution. No resolution switching.}
   \label{fig:cdn-constraint}
\end{center}
 \end{figure}

 \begin{figure}[t]
\begin{center}
\includegraphics[width=1\linewidth,draft=false]{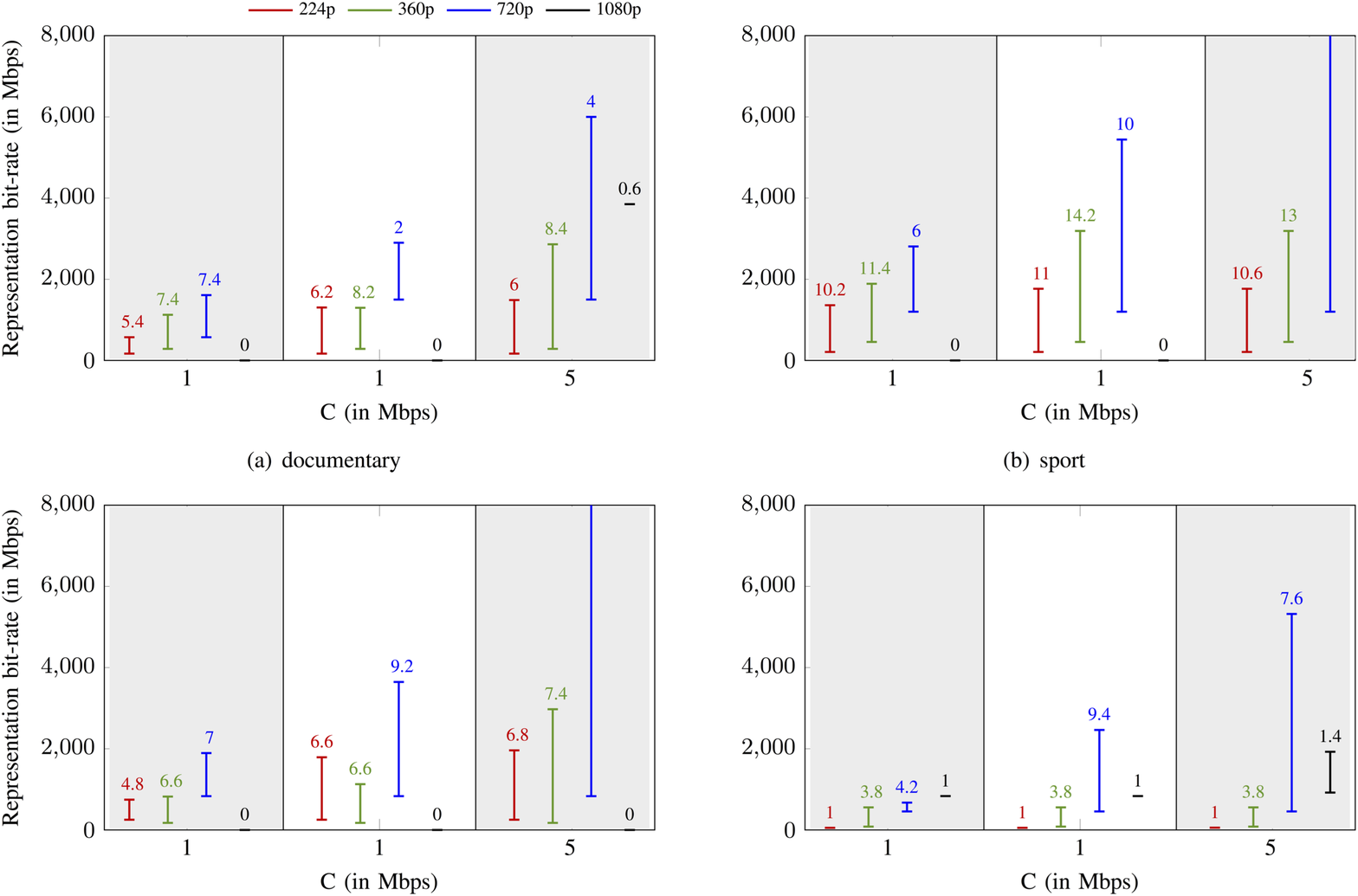}
    \caption{Range of representations when CDN capacity $C$ is limited. Three
  different CDN capacities are given. Bars are bounded, at the
  bottom (respectively top), by the average minimum (respectively
  maximum) value over 5 runs. The number over the bars indicates the
  average number of representations for the resolution.  Resolution switching.}
   \label{fig:constraint_switching}
\end{center}
 \end{figure}


Our first observation is that the higher the resolution, the broader the range 
of bit rates for the representations. Typically for the 1080p resolution, the 
bit rates ranges from $1,\!600\,kbps$ to almost $8,\!000\,kbps$. Such range is 
much larger than the one for the 224p resolution, from $200\,kbps$ to 
$2,\!300\,kbps$.  

Our second observation is that there exists a dense area of representations in 
the ``south west'' of every figure, meaning that there exist representations 
with the lowest possible rates in the optimal representations set, and that these 
representations are overall not accessed much. There are two reasons for such 
density in the low rates. First, the system has to ensure service for users 
connected by low capacity links (i.e., small values of $c_u$). It is thus 
necessary to have a representation at one of the lowest possible rates. Second, 
the gains in terms of QoE are usually large for low rates, so the encoding 
of a large number of representations at low rates is valuable because a small 
increase of the link capacity at the client side can result in a significant QoE 
gain.  In other words, at a given resolution, the distance between two 
consecutive representations in terms of encoding bit rate should be smaller for 
those representations with lower rates and higher for those ones with higher 
rates.

Our  third observation is that   no representation has a ``relative popularity" larger than three. Thus,   a constraint on the maximum number of users assigned to a 
representation is not necessary, although it would  be trivial to add it to the 
\ac{ILP} formulation.  Similar considerations can be derived from Fig.~\ref{fig:cloudpoint_switchi}, where resolution switching is allowed.

 

\vspace{0.35cm}
{\bf Guideline 4: How can we save CDN bandwidth still guaranteeing a good representations set?} \emph{To achieve low CDN budgets, the range of encoding rates used at each resolution should be narrow,    the number of representations per resolution should be limited, mainly for large resolutions, and at least
one representation at the minimum possible bit rate should be included in the optimal set for each resolution.  }
\vspace{0.15cm}

%

One of the major concerns of content providers is to reduce the costs of 
delivering video streams. Within this aim, we study the 
influence of the parameter $C$, the CDN capacity \emph{for each user}. The 
analysis of the  optimal representations sets aims at identifying ways to maintain a 
good average user satisfaction in under-provisioned configurations.



In Fig.~\ref{fig:cdn-constraint}, we focus on three critical CDN capacities: 
$C=1,\!000\,kbps$, $C=1,\!500\,kbps$, and $C=3,\!000\,kbps$. An average CDN 
budget  $C=1,\!000\,kbps$ represents a threshold value below which poor QoE 
(below $0.6$) values  are experienced on average for the requested 
representations. A budget of $C=3,\!000\,kbps$ is rather a threshold value above 
which an improvement of QoE is no more experienced. This means that in our setting, 
$C=3,\!000\,kbps$ leads the system to be not constrained by the CDN budget in 
Eq.~\eqref{eq:tot_capac}, achieving good QoE scores (above $0.9$). 
Finally, we considered an intermediate value $C=1,\!500\,kbps$, which should be 
enough to deliver a good quality of service to users (above $0.75$).   For each of these  
capacities, we provide the maximum and minimum  bit rates (averaged  over 5 runs) of the 
optimal representations sets. The number above the bar is the average number 
of representations per resolution and per video. The total number of 
representations $K$ is $100$ to be distributed among all videos and resolutions.

For a low capacity ($C=1,\!000\,kbps$), there are very few representations, 
only $65$ representations on average (evaluated by summing the number above the 
bar for all resolutions and videos in the $1,\!000\,kbps$ subplots) despite the 
maximum being $100$. The ranges of bit rates are very small as well. An 
efficient set of representations in such an under-provisioned context contains a 
few representations per resolution, at least one at the minimum possible bit 
rate. A similar trend is visible for $C=1,\!500\,kbps$. The number of 
representations increases, but the ranges of bit rates are still small. Note that similar trends are observed in Fig.~ $13$, where resolution switching is allowed at the decoder.

Finally, the scenario where $C=3,\!000\,kbps$ confirms our three first 
guidelines. The ranges of bit rates are larger for high resolutions, the number 
of representations depends on the videos and the number of representations is 
slightly higher for higher resolutions.

\section{Conclusions} 
\label{sec:concl} 

In this paper, we have proposed a new  optimization problem for the selection of 
the representations set that maximizes  the average satisfaction of users in 
adaptive streaming systems. We modeled this problem as an  integer linear 
program, whose  optimal solution can be computed by a generic solver.  The 
optimal set of representations is defined as the one that maximizes the users' 
satisfaction, given  information about users population, network dynamics, and  
video content.  We have conducted a  detailed numerical analysis of the 
performance of the optimal representations sets and the ones based on 
recommendations from system manufacturers and content providers.   We have also 
derived practical guidelines for  system engineers in charge of the encoding 
process in adaptive streaming  delivery systems.  Most of our study have 
considered dynamic network profiles for a given audience. As future works, we 
envision to extend our study to dynamic clients requests.  

This paper opens a large number of perspectives. It reveals the gap between 
existing recommendations and solutions that maximize the average user 
satisfaction. Although the representations sets can  severely impact the average 
QoE of users in adaptive streaming, this topic is  still overlooked in the 
literature.  We therefore outline   the importance of optimizing the 
representations sets in  today's   video delivery systems. We gather information 
from various engineers   and stakeholders to build a reasonable  model   in both 
  theoretical and practical contexts. The large number of parameters   to take 
into account when addressing optimization problems in this area however pose 
important  challenges. This paper is a first step toward a better understanding 
of the interaction and correlation between the numerous system parameters and 
the different blocks of the video delivery chain. It opens new  perspectives toward the design of processes  
that automatically set encoding parameters at the ingest server of content 
delivery architectures.

\appendix
\section{Q\lowercase{o}E Metric} 
In this section, we provide further details on  the user satisfaction function  of Eq.~\eqref{eq:QoE}, which is given by  
$$
    f_{uvrs} =1 - \left( m_{uvs} + \frac{n_{uvs}}{b_r+o_{uvs}}\right). 
$$
In Table~\ref{tab:test_QoE} we show   the parameters $m_{uvs}, n_{uvs},$ and $o_{uvs}$  
used in the  curve fitting process for each video $v$ and resolution $s$ to be 
displayed at size $s_u$.  

\begin{table}[t] 
\centering
\scriptsize
\begin{tabular}{c ccccc c ccccc c}
\toprule
\phantom{a} &\multicolumn{5}{c}{{Video: Rush Field Cuts}} & \phantom{a} & \multicolumn{5}{c}{{Video: Big Buck Bunny}} & \phantom{a}\\ 
\cmidrule{2-6}   \cmidrule{8-12}  
 &  Display & Res.  &       m       &       n       &  o    & &  
Display & Res. &       m       &       n       &  o    &\\
&  Size &   &   &    &     & &  
Size &  &  &    &    &\\
\cmidrule{2-6}    \cmidrule{8-12}   
& 224	& 224 	&   -0.10 &  188.63   & 196.92  			& &  224 & 224  &  -0.02 &   35.60  &  31.63  \\
& 224	&  360  	&   -0.04    & 167.48  &  62.29 		& &  224 & 360  &  0.11	 & 2.045   &  -87.70    \\
& 360	& 224 	&   0.04   &  219.79  	&    235.89     		&  &360 & 224 	& 0.05	 & 14.46  &  -60.65 \\ 
& 360	& 360 	&   -0.12   &  445.59 	 &    422.25 		&   &360   & 360  & -0.02 	 & 49.20  &  116.24   \\
& 360	& 720 	&   -0.06  &   339.13  	&    -164.01 		&  &360 &720 	& 0.04 	 & 23.97  &  -800.08  \\
& 720	& 360 	&   0.06  &   447.38  	&    426.25 		&  &720 & 360 	&  0.09	 & 23.37  &  22.26  \\
& 720	& 720 	&   -0.10  &   1348.64  	&    1574.48 		&  &720 & 720 	& -0.03	 & 166.45  &  -65.56    \\
& 720	& 1080  &   -0.03    &    852.28   &    262.06 		&  & 720 & 1080 & -0.01	&  80.94  &  -1156.78   \\
& 1080	 &  720  &   -0.03  &   1137.04    &   1025.20 		&  &1080 &	720 	&  -0.07 & 511.04  &  1834.94   \\
& 1080	 &  1080  &   -0.07    &   1548.17   &  1286.62 		&   &1080 &	1080 	&  -0.01  &  127.78  &  -523.06    \\
\midrule
\phantom{a} &\multicolumn{5}{c}{{Video: Snow Mountain}} & \phantom{a} & \multicolumn{5}{c}{{Video: Old Town Cross}} & \phantom{a}\\ 
\cmidrule{2-6}   \cmidrule{8-12}  
 &  Display & Res.  &       m       &       n       &  o    & &  
Display & Res.  &    m       &       n       &  o    &\\
 &  Size & &  &   &    & &  
Size &   & &  &   &\\
\cmidrule{2-6}    \cmidrule{8-12}   
& 224	& 224 	  &  -0.014 &   19.50 & 	  -68.49            & &  224           & 224 	   & -0.04   &   77.867   &   	  150.03    \\
& 224	& 360 	  &   0.001 &   21.32  &    -120.68 	& &  224	& 360 	   & 0.02  &    65.49   &    	  86.00     \\
& 360	& 224 	  &   0.09 &   25.49 	&   -55.62 	& &  360	& 224 	   &   0.07   &    112.80   &    	  243.34     \\
& 360	& 360 	  & -0.02  &   52.52  & 	  -105.32 	& &  360	& 360 	   & -0.04     &  136.26   &   	  259.10  \\
& 360	& 720 	  &  0.01 &   74.37  	&   -371.80 	& & 360	& 720 	   & -0.04   &    462.16   &   	  4214.38 \\
& 720	& 360 	  & 0.038  &   106.18  &    89.47 	& &  720	& 360 	   & 0.09     &  226.49   &   	  477.13  \\
& 720	& 720 	  & -0.018 & 187.43   &    -74.22 	& & 720	& 720 	   &  -0.01    &   119.49   &    	  -543.77 \\
& 720	& 1080   & 0.01  &  204.12  	&   -636.24  	& &  720	& 1080    & 0.04  &    148.76   &    	  -288.90  \\
& 1080	& 720 	  &  -0.04 &   414.67   &    704.83            & &  1080	& 720 	  &  -0.04  &    270.34   &    	  -61.45    \\
& 1080	& 1080   &  -0.03 &   372.06   	&   -165.76 	& &  1080	& 1080   &  0.02   &     148.38   &    	  -1498.73 \\
\bottomrule
\end{tabular}
\caption{Parameters of the satisfaction function model.}\label{tab:test_QoE}
\end{table}

\begin{figure}
  \centering
  \newcommand\largeur{0.5}
\newcommand\hei{4.5cm}

\subfigure[Display 224p]{\label{qoe-cartoon_224}
    \begin{tikzpicture}[scale=1]
     \begin{axis}[
       width=\largeur\columnwidth,
       height=\hei,
        ymin=0.6,   
        ymax=1,  
        xmin=0,
        xmax=9000,
        ylabel near ticks, xlabel near ticks,
        minor tick num = 1,
        xlabel={rate (in kbps)},
        ylabel={(1-VQM) normalized},
        enlargelimits=0.025,
        mark options = {solid},
        legend style={font=\scriptsize,legend pos=south east, legend
          columns = 1, draw=none},
        legend cell align=left
]

        \pgfplotstableread{QoE_resol/cartoon_224_224.csv}\loadedtable;
        \addplot[mark=o,only marks,green] table[x index=0, y index=1] {\loadedtable};
        \addplot[no marks,green] table[x index=0, y index=2] {\loadedtable};

        \pgfplotstableread{QoE_resol/cartoon_224_360.csv}\loadedtable;
        \addplot[mark=o,only marks,blue] table[x index=0, y index=1] {\loadedtable};
        \addplot[no marks,dotted,blue] table[x index=0, y index=2] {\loadedtable};

        \legend{,224p,,360p,}

      \end{axis}
    \end{tikzpicture}
}
\qquad
\subfigure[Display 360p]{\label{qoe-cartoon_360}
    \begin{tikzpicture}[scale=1]
     \begin{axis}[
       width=\largeur\columnwidth,
       height=\hei,
        ymin=0.6,   
        ymax=1,  
        xmin=0,
        xmax=9000,
        ylabel near ticks, xlabel near ticks,
        minor tick num = 1,
        xlabel={rate (in kbps)},
        enlargelimits=0.025,
        mark options = {solid},
        legend style={font=\scriptsize,legend pos=south east, legend
          columns = 1, draw=none},
        legend cell align=left
]
     
        \pgfplotstableread{QoE_resol/cartoon_360_224.csv}\loadedtable;
        \addplot[mark=o,only marks,green] table[x index=0, y index=1] {\loadedtable};
        \addplot[no marks,green] table[x index=0, y index=2] {\loadedtable};

        \pgfplotstableread{QoE_resol/cartoon_360_360.csv}\loadedtable;
        \addplot[mark=o,only marks,blue] table[x index=0, y index=1] {\loadedtable};
        \addplot[no marks,dotted,blue] table[x index=0, y index=2] {\loadedtable};

        \pgfplotstableread{QoE_resol/cartoon_360_720.csv}\loadedtable;
        \addplot[mark=o,only marks,magenta] table[x index=0, y index=1] {\loadedtable};
        \addplot[no marks,dashed,magenta] table[x index=0, y index=2] {\loadedtable};

        \legend{,224p,,360p,,720p}

      \end{axis}
    \end{tikzpicture}
}
\subfigure[Display 720p]{\label{qoe-cartoon_720}
    \begin{tikzpicture}[scale=1]
     \begin{axis}[
       width=\largeur\columnwidth,
       height=\hei,
        ymin=0.6,   
        ymax=1,  
        xmin=0,
        xmax=9000,
        ylabel near ticks, xlabel near ticks,
        minor tick num = 1,
        xlabel={rate (in kbps)},
        ylabel={(1-VQM) normalized},
        enlargelimits=0.025,
        mark options = {solid},
        legend style={font=\scriptsize,legend pos=south east, legend
          columns = 1, draw=none},
        legend cell align=left
]

        \pgfplotstableread{QoE_resol/cartoon_720_360.csv}\loadedtable;
        \addplot[mark=o,only marks,blue] table[x index=0, y index=1] {\loadedtable};
        \addplot[no marks,blue] table[x index=0, y index=2] {\loadedtable};

        \pgfplotstableread{QoE_resol/cartoon_720_720.csv}\loadedtable;
        \addplot[mark=o,only marks,magenta] table[x index=0, y index=1] {\loadedtable};
        \addplot[no marks,dotted,magenta] table[x index=0, y index=2] {\loadedtable};

        \pgfplotstableread{QoE_resol/cartoon_720_1080.csv}\loadedtable;
        \addplot[mark=o,only marks,black] table[x index=0, y index=1] {\loadedtable};
        \addplot[no marks,dashed,black] table[x index=0, y index=2] {\loadedtable};

        \legend{,360p,,720p,,1080p}

      \end{axis}
    \end{tikzpicture}
}
\qquad
\subfigure[Display 1080p]{\label{qoe-cartoon_1080}
    \begin{tikzpicture}[scale=1]
     \begin{axis}[
       width=\largeur\columnwidth,
       height=\hei,
        ymin=0.6,   
        ymax=1,  
        xmin=0,
        xmax=9000,
        ylabel near ticks, xlabel near ticks,
        minor tick num = 1,
        xlabel={rate (in kbps)},
        enlargelimits=0.025,
        mark options = {solid},
        legend style={font=\scriptsize,legend pos=south east, legend
          columns = 1, draw=none},
        legend cell align=left
]

        \pgfplotstableread{QoE_resol/cartoon_1080_720.csv}\loadedtable;
        \addplot[mark=o,only marks,magenta] table[x index=0, y index=1] {\loadedtable};
        \addplot[no marks,magenta] table[x index=0, y index=2] {\loadedtable};

        \pgfplotstableread{QoE_resol/cartoon_1080_1080.csv}\loadedtable;
        \addplot[mark=o,only marks,black] table[x index=0, y index=1] {\loadedtable};
        \addplot[no marks,dotted,black] table[x index=0, y index=2] {\loadedtable};

           \legend{,720p,,1080p}

      \end{axis}
    \end{tikzpicture}
}
\caption{Curve fitting for all the considered  display resolutions for cartoon video. Lines are
  real measures taken from the video while circles represent the model.}
  \label{fig:Curve_Fitting_cartoon}
\end{figure}
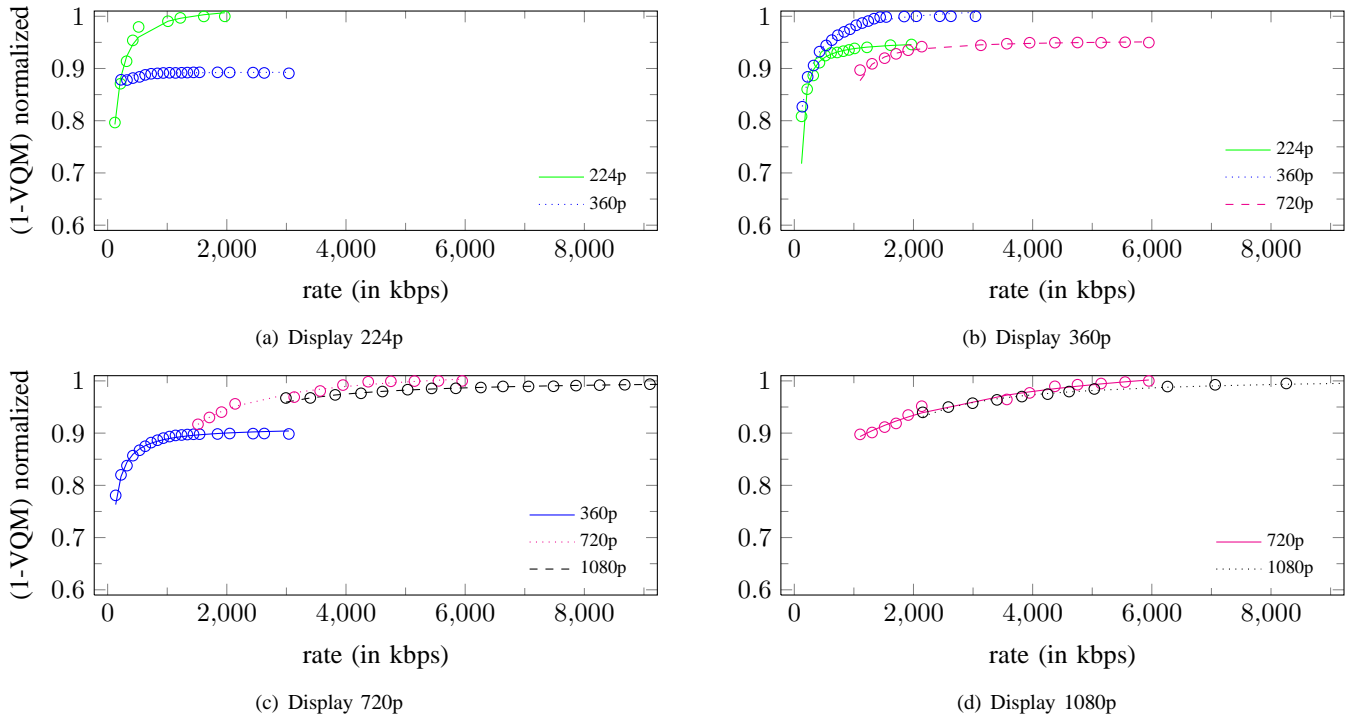

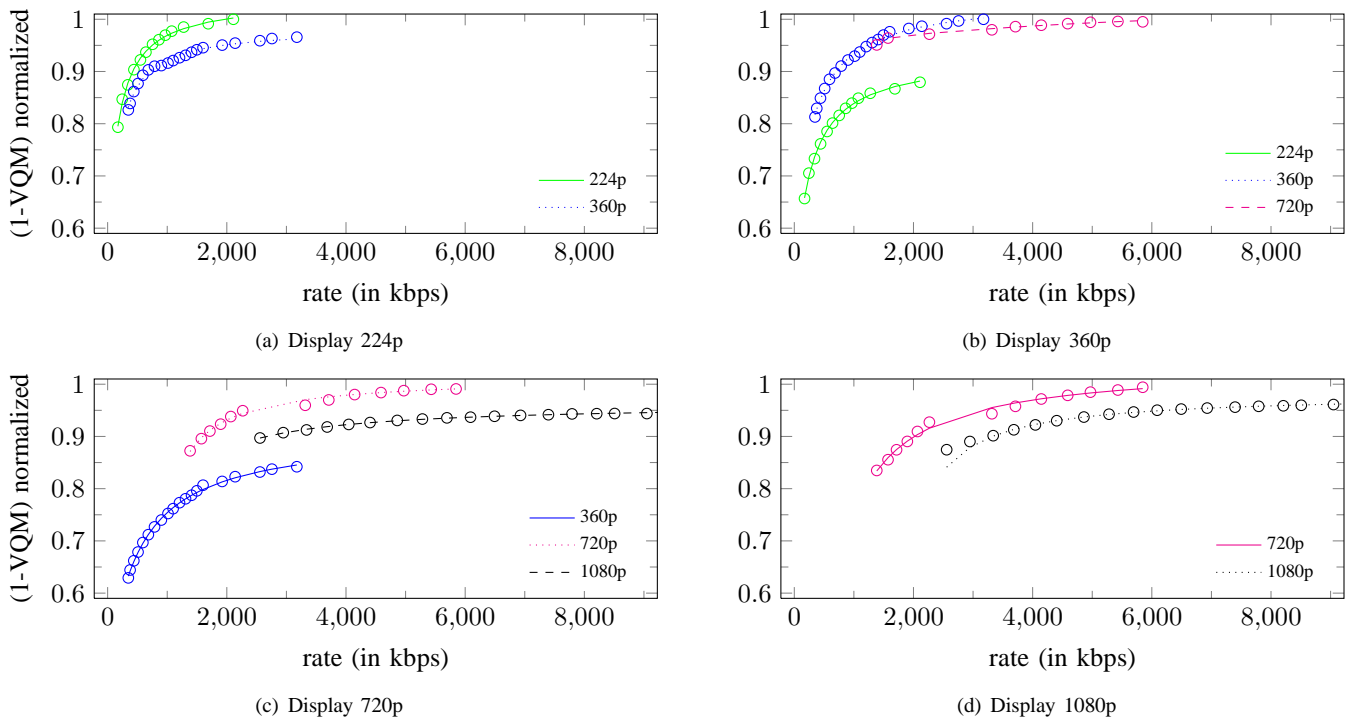
\begin{figure}
  \centering

  \newcommand\largeur{0.5}
\newcommand\hei{4.5cm}

\subfigure[Display 224p]{\label{qoe-video_224}
    \begin{tikzpicture}[scale=1]
     \begin{axis}[
       width=\largeur\columnwidth,
       height=\hei,
        ymin=0.6,   
        ymax=1,  
        xmin=0,
        xmax=9000,
        ylabel near ticks, xlabel near ticks,
        minor tick num = 1,
        xlabel={rate (in kbps)},
        ylabel={(1-VQM) normalized},
        enlargelimits=0.025,
        mark options = {solid},
        legend style={font=\scriptsize,legend pos=south east, legend
          columns = 1, draw=none},
        legend cell align=left
]

        \pgfplotstableread{QoE_resol/video_224_224.csv}\loadedtable;
        \addplot[mark=o,only marks,green] table[x index=0, y index=1] {\loadedtable};
        \addplot[no marks,green] table[x index=0, y index=2] {\loadedtable};

        \pgfplotstableread{QoE_resol/video_224_360.csv}\loadedtable;
        \addplot[mark=o,only marks,blue] table[x index=0, y index=1] {\loadedtable};
        \addplot[no marks,dotted,blue] table[x index=0, y index=2] {\loadedtable};

        \legend{,224p,,360p,}

      \end{axis}
    \end{tikzpicture}
}
\qquad
\subfigure[Display 360p]{\label{qoe-video_360}
    \begin{tikzpicture}[scale=1]
     \begin{axis}[
       width=\largeur\columnwidth,
       height=\hei,
        ymin=0.6,   
        ymax=1,  
        xmin=0,
        xmax=9000,
        ylabel near ticks, xlabel near ticks,
        minor tick num = 1,
        xlabel={rate (in kbps)},
        enlargelimits=0.025,
        mark options = {solid},
        legend style={font=\scriptsize,legend pos=south east, legend
          columns = 1, draw=none},
        legend cell align=left
]
     
        \pgfplotstableread{QoE_resol/video_360_224.csv}\loadedtable;
        \addplot[mark=o,only marks,green] table[x index=0, y index=1] {\loadedtable};
        \addplot[no marks,green] table[x index=0, y index=2] {\loadedtable};

        \pgfplotstableread{QoE_resol/video_360_360.csv}\loadedtable;
        \addplot[mark=o,only marks,blue] table[x index=0, y index=1] {\loadedtable};
        \addplot[no marks,dotted,blue] table[x index=0, y index=2] {\loadedtable};

        \pgfplotstableread{QoE_resol/video_360_720.csv}\loadedtable;
        \addplot[mark=o,only marks,magenta] table[x index=0, y index=1] {\loadedtable};
        \addplot[no marks,dashed,magenta] table[x index=0, y index=2] {\loadedtable};

        \legend{,224p,,360p,,720p}

      \end{axis}
    \end{tikzpicture}
}
\subfigure[Display 720p]{\label{qoe-video_720}
    \begin{tikzpicture}[scale=1]
     \begin{axis}[
       width=\largeur\columnwidth,
       height=\hei,
        ymin=0.6,   
        ymax=1,  
        xmin=0,
        xmax=9000,
        ylabel near ticks, xlabel near ticks,
        minor tick num = 1,
        xlabel={rate (in kbps)},
        ylabel={(1-VQM) normalized},
        enlargelimits=0.025,
        mark options = {solid},
        legend style={font=\scriptsize,legend pos=south east, legend
          columns = 1, draw=none},
        legend cell align=left
]

        \pgfplotstableread{QoE_resol/video_720_360.csv}\loadedtable;
        \addplot[mark=o,only marks,blue] table[x index=0, y index=1] {\loadedtable};
        \addplot[no marks,blue] table[x index=0, y index=2] {\loadedtable};

        \pgfplotstableread{QoE_resol/video_720_720.csv}\loadedtable;
        \addplot[mark=o,only marks,magenta] table[x index=0, y index=1] {\loadedtable};
        \addplot[no marks,dotted,magenta] table[x index=0, y index=2] {\loadedtable};

        \pgfplotstableread{QoE_resol/video_720_1080.csv}\loadedtable;
        \addplot[mark=o,only marks,black] table[x index=0, y index=1] {\loadedtable};
        \addplot[no marks,dashed,black] table[x index=0, y index=2] {\loadedtable};

        \legend{,360p,,720p,,1080p}

      \end{axis}
    \end{tikzpicture}
}
\qquad
\subfigure[Display 1080p]{\label{qoe-video_1080}
    \begin{tikzpicture}[scale=1]
     \begin{axis}[
       width=\largeur\columnwidth,
       height=\hei,
        ymin=0.6,   
        ymax=1,  
        xmin=0,
        xmax=9000,
        ylabel near ticks, xlabel near ticks,
        minor tick num = 1,
        xlabel={rate (in kbps)},
        enlargelimits=0.025,
        mark options = {solid},
        legend style={font=\scriptsize,legend pos=south east, legend
          columns = 1, draw=none},
        legend cell align=left
]

        \pgfplotstableread{QoE_resol/video_1080_720.csv}\loadedtable;
        \addplot[mark=o,only marks,magenta] table[x index=0, y index=1] {\loadedtable};
        \addplot[no marks,magenta] table[x index=0, y index=2] {\loadedtable};

        \pgfplotstableread{QoE_resol/video_1080_1080.csv}\loadedtable;
        \addplot[mark=o,only marks,black] table[x index=0, y index=1] {\loadedtable};
        \addplot[no marks,dotted,black] table[x index=0, y index=2] {\loadedtable};

           \legend{,720p,,1080p}

      \end{axis}
    \end{tikzpicture}
}
\caption{Curve fitting for all the considered  display resolutions for a generic movie channel. Lines are
  real measures taken from the video while circles represent the model.}
  \label{fig:Curve_Fitting_video}
\end{figure}

\begin{figure}
  \centering

  \newcommand\largeur{0.5}
\newcommand\hei{4.5cm}

\subfigure[Display 224p]{\label{qoe-documentary_224}
    \begin{tikzpicture}[scale=1]
     \begin{axis}[
       width=\largeur\columnwidth,
       height=\hei,
        ymin=0.6,   
        ymax=1,  
        xmin=0,
        xmax=9000,
        ylabel near ticks, xlabel near ticks,
        minor tick num = 1,
        xlabel={rate (in kbps)},
        ylabel={(1-VQM) normalized},
        enlargelimits=0.025,
        mark options = {solid},
        legend style={font=\scriptsize,legend pos=south east, legend
          columns = 1, draw=none},
        legend cell align=left
]

        \pgfplotstableread{QoE_resol/documentary_224_224.csv}\loadedtable;
        \addplot[mark=o,only marks,green] table[x index=0, y index=1] {\loadedtable};
        \addplot[no marks,green] table[x index=0, y index=2] {\loadedtable};

        \pgfplotstableread{QoE_resol/documentary_224_360.csv}\loadedtable;
        \addplot[mark=o,only marks,blue] table[x index=0, y index=1] {\loadedtable};
        \addplot[no marks,dotted,blue] table[x index=0, y index=2] {\loadedtable};

        \legend{,224p,,360p,}

      \end{axis}
    \end{tikzpicture}
}
\qquad
\subfigure[Display 360p]{\label{qoe-documentary_360}
    \begin{tikzpicture}[scale=1]
     \begin{axis}[
       width=\largeur\columnwidth,
       height=\hei,
        ymin=0.6,   
        ymax=1,  
        xmin=0,
        xmax=9000,
        ylabel near ticks, xlabel near ticks,
        minor tick num = 1,
        xlabel={rate (in kbps)},
        enlargelimits=0.025,
        mark options = {solid},
        legend style={font=\scriptsize,legend pos=south east, legend
          columns = 1, draw=none},
        legend cell align=left
]
     
        \pgfplotstableread{QoE_resol/documentary_360_224.csv}\loadedtable;
        \addplot[mark=o,only marks,green] table[x index=0, y index=1] {\loadedtable};
        \addplot[no marks,green] table[x index=0, y index=2] {\loadedtable};

        \pgfplotstableread{QoE_resol/documentary_360_360.csv}\loadedtable;
        \addplot[mark=o,only marks,blue] table[x index=0, y index=1] {\loadedtable};
        \addplot[no marks,dotted,blue] table[x index=0, y index=2] {\loadedtable};

        \pgfplotstableread{QoE_resol/documentary_360_720.csv}\loadedtable;
        \addplot[mark=o,only marks,magenta] table[x index=0, y index=1] {\loadedtable};
        \addplot[no marks,dashed,magenta] table[x index=0, y index=2] {\loadedtable};

        \legend{,224p,,360p,,720p}

      \end{axis}
    \end{tikzpicture}
}
\subfigure[Display 720p]{\label{qoe-documentary_720}
    \begin{tikzpicture}[scale=1]
     \begin{axis}[
       width=\largeur\columnwidth,
       height=\hei,
        ymin=0.6,   
        ymax=1,  
        xmin=0,
        xmax=9000,
        ylabel near ticks, xlabel near ticks,
        minor tick num = 1,
        xlabel={rate (in kbps)},
        ylabel={(1-VQM) normalized},
        enlargelimits=0.025,
        mark options = {solid},
        legend style={font=\scriptsize,legend pos=south east, legend
          columns = 1, draw=none},
        legend cell align=left
]

        \pgfplotstableread{QoE_resol/documentary_720_360.csv}\loadedtable;
        \addplot[mark=o,only marks,blue] table[x index=0, y index=1] {\loadedtable};
        \addplot[no marks,blue] table[x index=0, y index=2] {\loadedtable};

        \pgfplotstableread{QoE_resol/documentary_720_720.csv}\loadedtable;
        \addplot[mark=o,only marks,magenta] table[x index=0, y index=1] {\loadedtable};
        \addplot[no marks,dotted,magenta] table[x index=0, y index=2] {\loadedtable};

        \pgfplotstableread{QoE_resol/documentary_720_1080.csv}\loadedtable;
        \addplot[mark=o,only marks,black] table[x index=0, y index=1] {\loadedtable};
        \addplot[no marks,dashed,black] table[x index=0, y index=2] {\loadedtable};

        \legend{,360p,,720p,,1080p}

      \end{axis}
    \end{tikzpicture}
}
\qquad
\subfigure[Display 1080p]{\label{qoe-documentary_1080}
    \begin{tikzpicture}[scale=1]
     \begin{axis}[
       width=\largeur\columnwidth,
       height=\hei,
        ymin=0.6,   
        ymax=1,  
        xmin=0,
        xmax=9000,
        ylabel near ticks, xlabel near ticks,
        minor tick num = 1,
        xlabel={rate (in kbps)},
        enlargelimits=0.025,
        mark options = {solid},
        legend style={font=\scriptsize,legend pos=south east, legend
          columns = 1, draw=none},
        legend cell align=left
]

        \pgfplotstableread{QoE_resol/documentary_1080_720.csv}\loadedtable;
        \addplot[mark=o,only marks,magenta] table[x index=0, y index=1] {\loadedtable};
        \addplot[no marks,magenta] table[x index=0, y index=2] {\loadedtable};

        \pgfplotstableread{QoE_resol/documentary_1080_1080.csv}\loadedtable;
        \addplot[mark=o,only marks,black] table[x index=0, y index=1] {\loadedtable};
        \addplot[no marks,dotted,black] table[x index=0, y index=2] {\loadedtable};

           \legend{,720p,,1080p}

      \end{axis}
    \end{tikzpicture}
}
\caption{Curve fitting for all the considered  display resolutions for documentary video. Lines are
  real measures taken from the video while circles represent the model.}
  \label{fig:Curve_Fitting_documentary}
\end{figure}
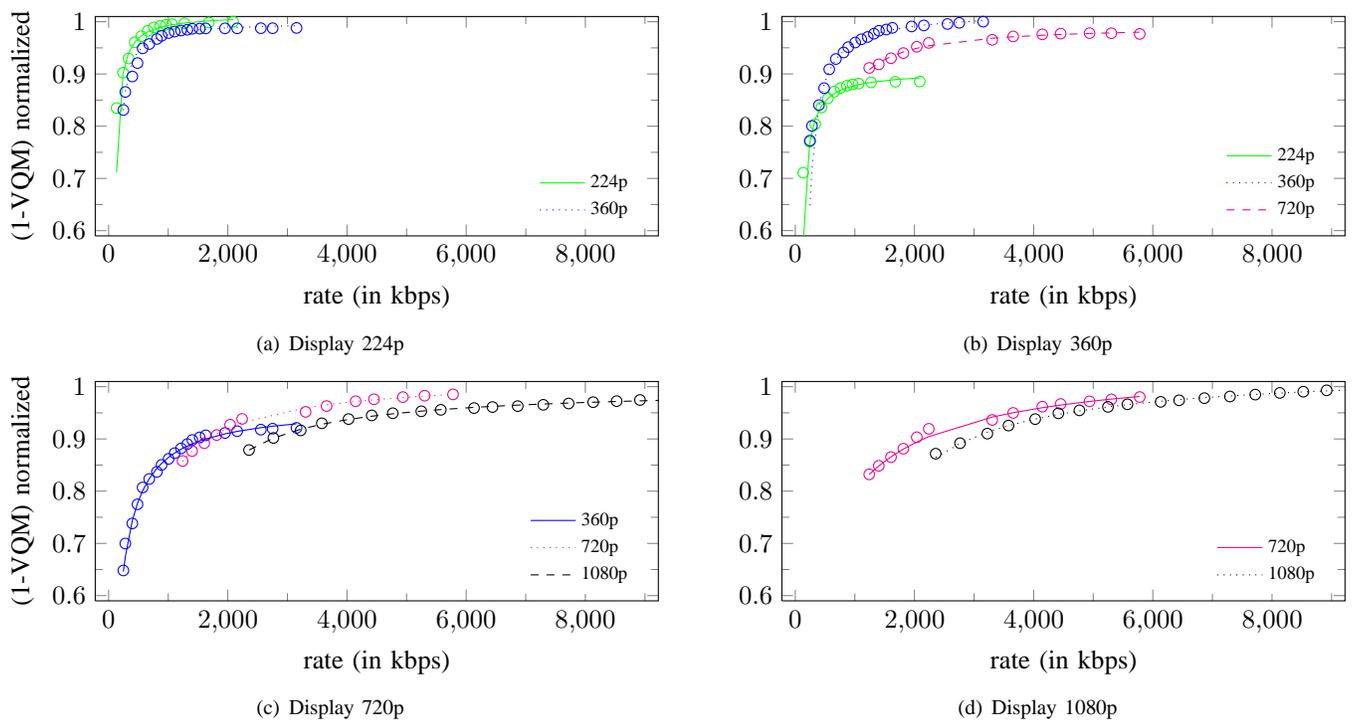

In Fig.~\ref{fig:Curve_Fitting_sport}, we have already compared the experienced QoE curves with the one from the  satisfaction curves evaluated from  Eq.~\eqref{eq:QoE}, for the sport video. 
In the following, we provide the user satisfaction curves for movie, cartoon, and 
documentary channel, respectively, in Fig. \ref{fig:Curve_Fitting_cartoon},  in 
Fig.  \ref{fig:Curve_Fitting_video}, and in  Fig. 
\ref{fig:Curve_Fitting_documentary}.  

It can be noticed that, for low display sizes (224p or 360p), the case with 
no up/down sampling (i.e., the case in which the display size is the same as the 
resolution size) is the one achieving the highest satisfaction. This is expected 
since  no additional artifacts are introduced due to spatial filtering. However, 
for larger display sizes it might be   more convenient to encode at an encoding 
resolution of 360p  and then perform the up-sampling rather than directly 
encode at  720p resolution. This can be observed in the sport and documentary 
channels. Similar also for the display size of 1080p.

\clearpage

\bibliographystyle{IEEEtran}
\bibliography{references_DASH} 
 
 
\end{document}